\documentclass[twocolumn,astrosymb,floatfix]{aastex631}

\graphicspath{{./}{figures/}}

\usepackage{savesym}
\savesymbol{tablenum}
\usepackage{siunitx}
\restoresymbol{SIX}{tablenum}

\usepackage{bm}
\usepackage{amsmath}
\usepackage{mathtools}
\usepackage{threeparttable}
\usepackage{booktabs}
\usepackage{makecell}
\usepackage[caption=false]{subfig}

\shorttitle{Three-Dimensional Simulations of Massive Stars: II. Age Dependence}
\shortauthors{Vanon et al.}

\begin{document}

\title{Three-Dimensional Simulations of Massive Stars: II. Age Dependence}
\author[0000-0001-5747-8476]{R. Vanon}
\altaffiliation{Current address: Department of Applied Mathematics, School of Mathematics, University of Leeds, Leeds, LS2 9JT, UK}
\affiliation{School of Mathematics, Statistics and Physics, Newcastle University, Newcastle upon Tyne, NE1 7RU, UK}

\author[0000-0001-7019-9578]{P.~V.~F. Edelmann}
\affiliation{Computer, Computational and Statistical Sciences (CCS) Division and Center for Theoretical Astrophysics (CTA), Los Alamos National Laboratory, Los Alamos, NM 87545, USA}

\author{R.~P. Ratnasingam}
\affiliation{Nicolaus Copernicus Astronomical Center, Polish Academy of Sciences, Poland}

\author{A. Varghese}
\affiliation{School of Mathematics, Statistics and Physics, Newcastle University, Newcastle upon Tyne, NE1 7RU, UK}

\author{T.~M. Rogers}
\affiliation{School of Mathematics, Statistics and Physics, Newcastle University, Newcastle upon Tyne, NE1 7RU, UK}
\affiliation{Planetary Science Institute, Tucson, AZ 85721, USA}

\begin{abstract}
    We present 3D full star simulations, reaching up to 90$\%$ of the total stellar radius, for three $7M_\odot$ stars of different ages (ZAMS, midMS and TAMS). A comparison with several theoretical prescriptions shows the generation spectra for all three ages are dominated by convective plumes. Two distinct overshooting layers are observed, with most plumes stopped within the layer situated directly above the convective boundary (CB); overshooting to the second, deeper layer becomes progressively more infrequent with increasing stellar age. Internal gravity wave (IGW) propagation is significantly impacted in the midMS and TAMS models as a result of some IGWs getting trapped within their Brunt--V\"{a}is\"{a}l\"{a} frequency spikes. A fundamental change in the wave structure across radius is also observed, driven by the effect of density stratification on IGW propagation causing waves to become evanescent within the radiative zone, with older stars being affected more strongly. We find that the steepness of the frequency spectrum at the surface increases from ZAMS to the older models, with older stars also showing more modes in their spectra. 
\end{abstract}

\section{Introduction} \label{sec:intro}
Internal Gravity Waves (IGWs) are waves that propagate in stably stratified fluids for which the restoring force is buoyancy. These waves are generated when a disturbance is present within the stratified medium and they have been found to play an important role in the large-scale dynamics in a variety of systems, including atmospheres, oceans and stars.
In Earth's atmosphere, for example, IGWs are instrumental in driving the large-scale winds known as quasi-biennial oscillations (QBOs) in both the equatorial stratosphere and the mesosphere \citep{Baldwinetal2001}.

In stars, IGWs are generated by any disturbance to the radiation zone, with the most common source being stochastically generated perturbations at the convective-radiative interface caused by either overshooting convective plumes \citep{Townsend1966,MontalbanSchatzman2000,RogersMacGregor2011} or Reynolds stresses of convective eddies \citep{Lighthill1952, GoldreichKumar1990, Kumaretal1999, LecoanetQuataert2013}. Furthermore, IGWs can be generated through tidal forcing by a companion star or planet \citep[e.g.,][]{Zahn1975, Fuller2017}. 
Once generated, IGWs propagate within the radiation zone away from the convective boundary (CB), resulting in an inward propagation in stars with convective envelopes ($M\lesssim 1.2M_\odot$) and an outward propagation in stars with convective cores ($M\gtrsim 1.2M_\odot$). IGWs are likely to be particularly important in the latter case due to the waves travelling along a negative density gradient which allows a significant amount of wave amplification despite radiative damping.

Significant discrepancies between observational constraints and stellar models have led to the search for mechanisms capable of efficiently transporting angular momentum and mixing chemical species through stellar radiative zones \citep[see e.g.,][]{Aertsetal2019}. These divergences have remained despite stellar models adopting rotation-induced angular momentum transport mechanisms such as meridional circulation and shear-induced turbulence \citep{Eggenbergeretal2012, Ceillieretal2013, Eggenbergeretal2019}. Additional transport mechanisms are therefore being investigated, such as magnetism \citep{Spruit2002, Hegeretal2005,Rudigeretal2015}, mixed modes \citep{Belkacemetal2015a, Belkacemetal2015b} and waves \citep{Press1981, TalonCharbonnel2005, Rogersetal2013}.  

IGWs are believed to play an important role in stellar dynamics thanks to their ability to transport angular momentum and to contribute to chemical mixing. For this reason, many studies have looked at the feasibility of IGWs as a possible way to reconcile stellar models and poorly explained observations in stellar astrophysics, such as stellar cores rotating in the opposite direction as their envelopes \citep{Trianaetal2015,Rogers2015}, the internal stellar rotation structure \citep{Becketal2012,Aertsetal2017,VanReethetal2018}, the enhanced mass loss required to explain certain core collapse supernovae \citep{QuataertShiode2012}, and Li depletion in F stars \citep{GarciaLopezSpruit1991} or in the Sun \citep{Schatzman1993,Montalban1994,TalonCharbonnel2005}. IGWs are also a plausible candidate for the stochastic low-frequency photometric variations observed in massive stars, with observed amplitude spectra at the stellar surface comparing well to those produced by numerical simulations \citep{AertsRogers2015,Bowmanetal2019,Bowmanetal2019b,Bowmanetal2020}; the origin of such variability is however highly debated, with sub-surface convection \citep{Cantielloetal2009, Lecoanet2019} and stellar wind instabilities \citep{KrtickaFeldmeier2021} also having been suggested to explain the phenomenon. 

The importance of IGWs in stellar internal dynamics and their potential involvement in the phenomena described above strongly depends on the properties of the IGW generation spectrum at the CB, which is unconstrained by observations. 
Most of the studies develop theories for the IGW generation spectrum and the IGW amplitudes based on the assumption that the generation spectrum is exclusively driven by either plume incursion or by convective eddies.
For the former, \cite{Townsend1966} was the first to analyse a plume-generated spectrum within the context of the Earth's atmosphere, and the analysis was subsequently applied to stellar systems by \cite{MontalbanSchatzman2000}. More recently, \cite{Pinconetal2016} and \cite{Pinconetal2017} developed semi-analytical models for the IGW flux triggered by penetrating plumes in solar-like and sub-giants stars, respectively. All works modelling plume-induced IGWs yield similar forms for the expected spectrum, which is found to be proportional to $\exp \left[- \left(f/f_\mathrm{pl}\right)^2\right]$, where $f$ and $f_\mathrm{pl}$ are the wave and plume frequencies, respectively. These spectra favour very low frequencies, with their steepness being dictated by the timescale of the plume, with shorter timescales resulting in flatter spectra and gentler slopes. 
On the other hand, works such as \cite{Lighthill1952, GoldreichKumar1990, Kumaretal1999, LecoanetQuataert2013} assumed Reynolds stresses caused by turbulent eddies to be the dominant driving mechanism for IGWs. This resulted in spectra with steep power law behaviour, i.e. $\propto f^{-\alpha}$, with the exponent $\alpha$ being dependent on the properties of the Brunt--V\"{a}is\"{a}l\"{a} frequency profile at the convective-radiative boundary. Most of the energy in these spectra is concentrated at the turnover frequency, with significant drops in amplitude away from it.

In reality, both mechanisms would contribute towards driving IGWs, meaning the resulting generation spectrum is likely to be a combination of the two forms discussed above. 
This outlines the importance of multi-dimensional hydrodynamical numerical simulations of stellar interiors -- where convective motions, IGW generation and propagation occur organically as a result of solving the fluid equations -- in understanding the role of IGWs in stellar dynamics and their expected signatures in observations. Despite numerical simulations being forced to implement kinematic viscosities and thermal diffusivities which are much larger than stellar values due to numerical constraints, they still provide an important benchmark for theoretical predictions and grant a fundamental link between models and observations. 

For example, it was found that IGWs are not able to explain the solar Li depletion \citep{Rogersetal2006}, while they can only partly account for the uniform rotation of the solar radiative interior \citep{RogersGlatzmaier2005, Denissenkovetal2008}. Furthermore, 2D hydrodynamical simulations by \cite{Rogersetal2013} of massive stars, focusing on IGW generation at the CB, found generation spectra with much shallower slopes than theoretically predicted and clearly exhibiting a broken power law behaviour, suggesting both IGW driving mechanisms (plumes and eddies) being relevant. More recently, research on IGW non-linearity in stellar radiative envelopes found that the amount of wave breaking is highly sensitive to their generation spectrum, with simulation spectra \citep[such as][]{Rogersetal2013} leading to increased wave breaking compared to theoretical models, therefore resulting in an enhanced angular momentum transport and chemical mixing via IGWs \citep{Ratnasingametal2019}.
Lastly, 3D simulations of a 3M$_\odot$ Zero Age Main Sequence (ZAMS) star by \cite{Edelmannetal2019}, which consider up to 90$\%$ of the whole star, confirmed the findings of IGW--dominated surface spectra and of broken power law propagation spectra found by 2D simulations \citep[such as][]{Rogersetal2013,Ratnasingametal2020, Horstetal2020}. This appears to validate the effectiveness of the latter at studying IGWs in stellar settings, as well as being consistent with recent asteroseismological observations of stochastic low-frequency variations \cite[e.g.][]{AertsRogers2015,Bowmanetal2019b}, although their origin remains debated \citep[e.g.]{Lecoanet2019}. 

The rest of the paper is structured as follows: Section~\ref{sec:code} introduces the hydrodynamical equations being solved in our 3D simulations as well as their discretization within the pseudo-spectral framework of the code; Section~\ref{sec:sims} describes the MESA models which are used as reference states for our simulations, and identifies how numerical restrictions were managed. In Sections~\ref{sec:results-generation}--\ref{sec:results-dominance} we analyze the generation spectra from our simulations and compare them to results from the literature to assess whether they are dominated by overshooting plumes, whose properties within our simulations are analysed in Sec.~\ref{sec:results-plumes}, or convective eddies. 
Section~\ref{sec:results-propagation} examines how IGW propagation throughout the radiation zone is affected by stellar age and discusses the causes behind these changes; 
the results section concludes with Section~\ref{sec:results-angularmomentum}, which examines the interplay between the Reynolds and viscous fluxes in the IGW-mediated angular momentum transfer in our simulations. Finally, we discuss the results of our work in Section~\ref{sec:conclusions}.

\section{Numerical method} \label{sec:code}
This work represents an extension of the simulations presented in \cite{Edelmannetal2019} and therefore employs the same 3D hydrodynamical numerical method based on the anelastic approximation with the horizontal component of the equations being discretized in terms of spherical harmonics. Below we give a condensed overview of the numerical method and its properties, but we refer the interested reader to \cite{Edelmannetal2019} for more detailed information, such as the resolution dependence or the parallelization efficiency.  

Considering an initial reference state, which is obtained from a stellar evolution model in hydrostatic equilibrium and is denoted by a bar (e.g. $\overline{T}$), we solve the following set of equations for the deviation from such a state

\begin{equation} \label{eq:continuity}
	\nabla \cdot \overline{\rho} \bm{v} = 0 ,
\end{equation} \vspace*{-20pt}

\begin{equation} \label{eq:Navier-Stokes}\begin{multlined}
	\frac{\partial \bm{v}}{\partial t} = - \left(\bm{v} \cdot \nabla \right) \bm{v} - \nabla P - C \, \overline{\bm{g}} + 2\left( \bm{v} \times \bm{\Omega} \right) \\ 
    + \nu \left( \nabla^2 \bm{v} + \frac{1}{3} \nabla \left(\nabla \cdot \bm{v} \right) \right) ,
\end{multlined}  
\end{equation}
\begin{equation}
\begin{multlined} \label{eq:temperature}
	\frac{\partial T}{\partial t} = -\left( \bm{v} \cdot \nabla \right) T + \left( \gamma-1\right) T h_\rho v_r \\ 
	- v_r \left( \frac{\mathrm{d} \overline{T}}{\mathrm{d} r} - \left( \gamma-1 \right) \overline{T} h_\rho \right) + \frac{\overline{Q}}{c_v \overline{\rho}} \\
	+ \frac{1}{c_v \overline{\rho}} \nabla \cdot \left( c_p \overline{\kappa} \overline{\rho} \nabla T\right) + \frac{1}{c_v \overline{\rho}} \nabla \cdot \left( c_p \overline{\kappa \rho} \nabla \overline{T} \right) .
\end{multlined}\end{equation}
where $\overline{\rho}$ is the reference state density, $\bm{v}$ is the 3D velocity, $\overline{\bm{g}}=\overline{g} \hat{\bm{r}}$ the background gravitational acceleration, $\nu$ and $\kappa$ are the kinematic viscosity and the thermal diffusivity, $\bm{\Omega}=\Omega \hat{\bm{z}}$ is the star's angular velocity, $T$ is the temperature deviation from the background temperature $\overline{T}$, $\gamma$ is the adiabatic index, $h_\rho$ is the inverse density scale height, $\overline{Q}$ is the rate at which energy is released and, lastly, $c_v$ is the specific heat capacity at constant density. 
Self-gravitational perturbations have also been considered, at no additional computational cost, by defining the reduced pressure $P$ and the co-density $C$ \citep{BraginskyRoberts1995, RogersGlatzmaier2005} as

\begin{equation} \label{eq:red-pressure}
	 P = p/\overline{\rho} + \Phi ,
\end{equation} 
\begin{equation} \label{eq:co-density}
	 C = - \frac{1}{\overline{T}} \left( T + \frac{1}{\overline{g \rho}} \frac{\partial \overline{T}}{\partial z} p \right) .
\end{equation}

As explained in more detail in \cite{Edelmannetal2019}, the code used here -- which employs a spherical coordinate system with radius $r$, colatitude $\theta$ and longitude $\phi$ -- is pseudo-spectral in nature and chooses a numerical method which is similar to \cite{Glatzmaier1984} and to the ASH code \citep{Cluneetal1999}. The main difference from these codes is that we use a finite-difference discretization of the radial derivatives, which allows us to adapt the radial grid spacing to the underlying stellar model.
The code solves for four unknowns: temperature $T$, reduced pressure $P$ and the poloidal and toroidal stream function components, $W$ and $Z$, which replace the momentum density through

\begin{equation} \label{eq:massflux-streamfunction}
	\overline{\rho} \bm{v} = \nabla \times \nabla \times W \hat{\bm{r}} + \nabla \times Z \hat{\bm{r}},
\end{equation}
allowing us to implicitly fulfill Eq.~\ref{eq:continuity}. These unknowns are then expanded by means of a spherical harmonics decomposition, resulting in the following for the reduced pressure $P$

\begin{equation}
	P \left(r,\theta,\phi,t\right) = \sum_{m=-m_{\mathrm{max}}}^{m_{\mathrm{max}}} \, \, \sum_{l=\lvert m \rvert}^{l_{\mathrm{max}}} P_{l,m}\left(r,t\right) Y_{l,m} \left(\theta,\phi\right),
\end{equation}
and similarly for the remaining unknowns. Here, $Y_{l,m}$ are the spherical harmonics with degree $l$ and azimuthal order $m$, and $P_{l,m}$ are the reduced pressure complex coefficients, which have a radial dependency. This decomposition allows for an inexpensive computation of the horizontal derivatives, as well as avoiding singularities at the stellar poles. 
Lastly, non-linear terms are dealt with using the explicit Adams--Bashforth method, while linear terms are calculated using the implicit Crank--Nicolson method which avoids the stringent Courant--Friedrichs--Lewy (CFL) condition.

\begin{figure*}
	\centering
	\subfloat{
    	\includegraphics[width=\columnwidth]{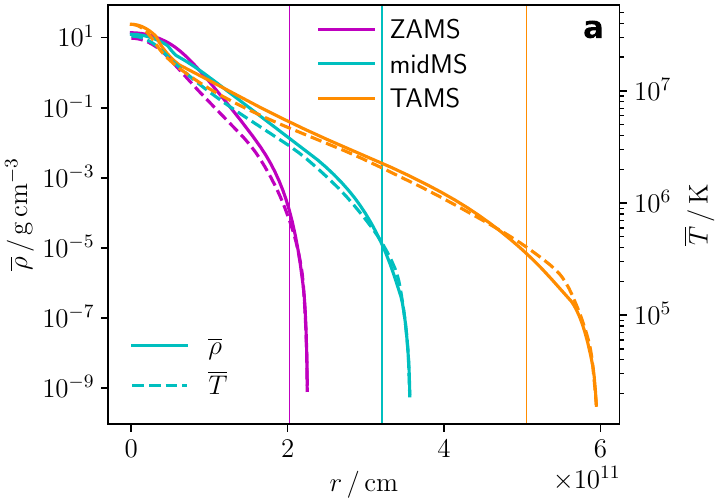}
        \label{fig:mesa_models}
        }
    \hspace*{2em}
    \subfloat{
        \includegraphics[width=\columnwidth]{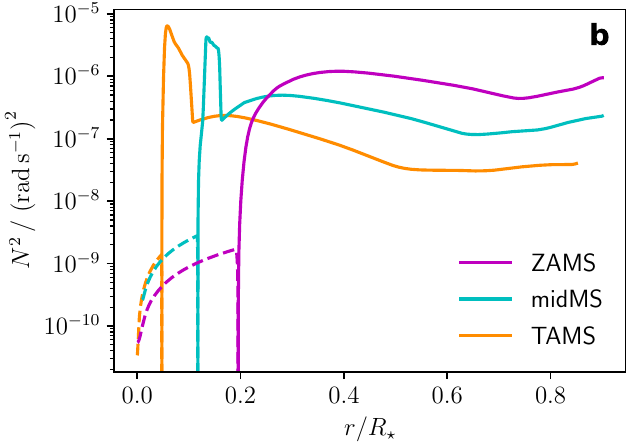}
	    \label{fig:sims_N2}
    }
    
	\caption{Radial profiles for \textbf{\textit{(a)}} density (full lines) and temperature (dashed) from the MESA reference states, and for \textbf{\textit{(b)}} the Brunt--V\"{a}is\"{a}l\"{a} frequency squared $N^2$ used in our three models: ZAMS (magenta), midMS (cyan), TAMS (orange). The colour-coded vertical lines used in \textit{(a)} mark the extent of the respective simulation domains, while the dashed portions of the $N^2$ profiles in \textit{(b)} indicate convectively unstable regions.}
\end{figure*}


\section{Simulation Setup} \label{sec:sims}
As mentioned at the beginning of Section~\ref{sec:code} we expand on the work carried out by \cite{Edelmannetal2019}, which focused on 3D hydrodynamical numerical simulations of a 3M$_\odot$ ZAMS star. We grow the sample of studied models by considering three 7M$_\odot$ stars at  different evolutionary stages: ZAMS, midMS and TAMS (Terminal Age Main Sequence); here ZAMS, midMS and TAMS are defined by a hydrogen mass fraction in the stellar core of $X_c=0.70$, $X_c=0.35$ and $X_c=0.011$, respectively. All models are uniformly rotating, with rotational rates ranging from $1.5\%$ to $15\%$ of the critical (or break-up) velocity $\Omega_\mathrm{crit} = \sqrt{GM_\star/R^3_\star}$. The metallicity is fixed at $Z=0.02$ for all runs.

\begin{table*}[ht!]
\makebox[\textwidth][c]{
\begin{threeparttable}
    \caption{Properties of 3D simulations.}
    \label{table:models}
    \begin{tabular*}{1.1\textwidth}{@{\extracolsep{\fill}}l|cccccccccc}
        \toprule \toprule

        \textbf{Model}\tnote{a} & $\bm{N_\phi}$, $\bm{N_\theta}$\tnote{b} & $\bm{N_r}$, $\bm{N_{r,CZ}}$\tnote{c} & $\bm{X_c}$\tnote{d} & $\bm{R_\star/\mathrm{cm}}$\tnote{e} & $\bm{\overline{\Omega}/10^{-6} \mathrm{rad \, s^{-1}}}$\tnote{f} & $\bm{\nu / 10^{13} \, \mathrm{cm^2 \, s^{-1}}}$ & $\bm{\kappa/\kappa_\star}$ & $\bm{\mathrm{Re}}$ & $\bm{\mathrm{Pr}}$ & $t_\mathrm{sim}/\mathrm{d}\tnote{h}$ \\ \midrule
            
        7ZR10 & 256, 128 & 1512, 400 & 0.70 & $2.25\times 10^{11}$ & 10 
        & $4.2,10.2$\tnote{g} & $3.0 \times 10^4,35$\tnote{g} & 510 & $5.6,0.40$\tnote{g} & $116.6$ \\ 

        7mR10 & 256, 128 & 2016, 629 & 0.35 & $3.56\times 10^{11}$ & 10 
        & $3.1,5.3$\tnote{g} & $9.5\times 10^{3},20$\tnote{g} & 735 & $6.3,0.035$\tnote{g} & $102.8$\\ 	
    
        7TR10 & 256, 128 & 2016, 400 & 0.010 & $5.95\times 10^{11}$ & 10 
        & $2.3,14.1$\tnote{g} & $3.0 \times 10^3,3$\tnote{g} & 530 & $15.1,0.27$\tnote{g} & $111.8$\\ 	

        
        7TR1 & 256, 128 & 2016, 400 & 0.010 & $5.95\times 10^{11}$ & 1 
        & $2.0,6.3$\tnote{g} & $5.7 \times 10^3,15$\tnote{g} & 655 & $6.8,0.024$\tnote{g} & $113.9$\\ 	
        
        \bottomrule \bottomrule
    \end{tabular*} 
    \begin{tablenotes}
    \item[a] Model name tag composed by: stellar mass (in units of $M_\star/M_\odot$), stellar age ($\mathrm{Z=ZAMS, \, m=midMS, \, T=TAMS}$) and rotation rate $R$ in units of $10^{-6} \, \mathrm{rad \, s^{-1}}$.
    \item[b] Real space resolution in longitude $N_\phi$ and latitude $N_\theta$
    \item[c] Total radial resolution $N_r$ and radial resolution in the convection zone $N_{r,CZ}$
    \item[d] Hydrogen mass fraction in the stellar core
    \item[e] Stellar radius
    \item[f] Background uniform rotation rate
    \item[g] Values averaged over the simulation: the first value applies to the CZ, the second to the domain's surface
    \item[h] Physical run time of the simulation
    \end{tablenotes}
    \end{threeparttable}
}
\end{table*}

The system of equations described in Section~\ref{sec:code} calculates the perturbation from a spherically symmetric reference state, which was obtained from the MESA (Modules for Experiments in Stellar Astrophysics) stellar evolution code \citep{Paxtonetal2011,Paxtonetal2013,Paxtonetal2015,Paxtonetal2018}. The radial profiles of temperature, density and gravity from the MESA reference states are interpolated onto the 3D spherical system of coordinates considered, with the convective cores possessing a finer resolution compared to their respective radiation zones to properly capture the turbulent nature of convection. Figure~\ref{fig:mesa_models} shows the density (full lines) and temperature (dashed lines) profiles of the reference states for the three models as a function of radius, highlighting the difference in size between the three stellar ages. The colour-coded vertical lines mark the radial extent of the 3D simulations, with both ZAMS and midMS stretching up to $90\%$, while for the larger TAMS we only consider $85\%$ of its radial domain. These domains result in all three models having densities between $\sim 10^{-4}$ and $\sim 10^{-5} \, \mathrm{g \, cm^{-3}}$ at their outer boundary.

In our simulations, we set convective/radiative regions by setting the super/sub-adiabatic temperature gradients defined as: 
\begin{equation}
- \left[\frac{\mathrm{d} \overline{T}}{\mathrm{d} r} - \left(\gamma - 1\right) \overline{T} h_\rho\right],
\end{equation}
In order to set this value, and to incorporate the stabilising effect of the chemical gradient left over from hydrogen burning as the star ages, we set this super/sub-adiabaticity using the Brunt--V\"{a}is\"{a}l\"{a} frequency (BVF) profiles from MESA: 
\begin{equation}
    N^2 = \frac{\overline{g}}{\gamma \overline{T}} \left( \frac{\mathrm{d}\overline{T}}{\mathrm{d}r} - (\gamma-1) \overline{T} h_\rho \right)
\end{equation}

With the two clearly related such that 

\begin{equation}
-\left[\frac{\mathrm{d}\overline{T}}{\mathrm{d} r}-\left(\gamma -1\right) \overline{T} h_\rho\right]=-N^2\frac{\gamma\overline{T}}{\overline{g}}
\end{equation}

Unfortunately, MESA profiles for super-adiabaticity are noisy, often yielding negative values, hence within the convection zones we simply set a constant super-adiabatic value of $10^{-6}$. It is worth noting that the literature has not come to a consensus on what represents a realistic super-adiabaticity within stellar convection zones.  
Regardless, given the code allows thermal transport, this value only directly influences the initial transient amplitude; once convection sets in, the adiabaticity in the CZ is determined by its effective value, which is subject to changes in the flow and heat transport, and therefore the initial value should have no effect on the long-term steady state properties.

Figure~\ref{fig:sims_N2} shows the profiles of the Brunt--V\"{a}is\"{a}l\"{a} frequency (BVF) squared $N^2$, defined as in Equation (9) used in the simulations for all three stellar ages as a function of the fractional stellar radius $r/R_\star$. The Brunt--V\"{a}is\"{a}l\"{a} frequency governs the propagation of IGWs, with only waves having a frequency $\omega$ smaller than $N$ ($\omega=2\pi f < N$) being able to propagate, although the stellar rotation rate $\Omega$ provides a lower limit on the available range of frequencies. 
Figure~\ref{fig:sims_N2} clearly illustrates the shrinking of the convective stellar core ($N^2$ values within convective zones are represented by dashed lines) during the stellar lifetime which leaves behind a steep compositional gradient, creating an increasingly prominent spike in the BVF. These BVF spikes have a significant effect on the interior dynamics of a star by trapping high frequency IGWs. Trapping in the spike occurs when high frequency waves have $\omega < N$ within the spike, but $\omega > N$ outside and hence are reflected, setting up standing modes within the spike, but preventing propagation in the bulk of the radiative zone. The IGW trapping becomes particularly substantial for the TAMS model, with waves larger than $\sim 70 \si{\micro Hz}$ being trapped. 

\begin{figure}[t] 
    \includegraphics[width=\columnwidth]{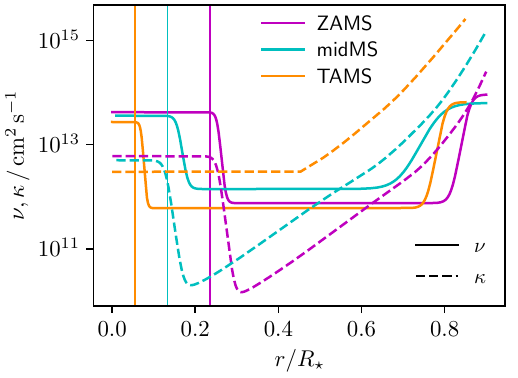}
    \caption{Radial profiles for the viscosity $\nu$ (full lines) and thermal diffusivity $\kappa$ (dashed lines) for our simulations. The colour-coded vertical lines indicate the size of the respective convective zones.}
	\label{fig:model-ZAMS-nu-kappa}
\end{figure}

The profiles for $\nu$ (full lines) and $\kappa$ (dashed lines) used in our simulations are shown in Fig.~\ref{fig:model-ZAMS-nu-kappa}. The thermal diffusivity profiles are kept constant within the convection zones (indicated by colour-coded vertical lines) and in the radiation zone they are smoothed down to reach the stellar profile multiplied by a constant factor (as indicated in Table~\ref{table:models}, e.g. $35\kappa_\star$ for the ZAMS run). TAMS was handled differently due to the sharp transition at the convective-radiative interface; here the constant $\kappa$ value is instead extended until about halfway across the RZ. The profiles for $\nu$ show a similarly constant values in the CZ and lower values in the radiation zone.  Increased $\nu$ near the simulated surface were needed to quench Kelvin-Helmholtz instabilities occurring because of the low densities there. The profiles of $\kappa$ and $\nu$ chosen represent a compromise between ensuring numerical stability, maximising convective turbulence and minimizing wave damping.  Although seemingly arbitrary, these diffusivities are "eddy/turbulent" diffusivities and hence, physically, would be larger in convective regions than in radiative regions.


Table~\ref{table:models} summarises the properties of our runs, including values for the \textit{Reynolds number} (in the convection zone):
\begin{equation} \label{eq:reynolds}
    \mathrm{Re} = \frac{v_\mathrm{rms}^{\mathrm{CZ}}D}{\nu},
\end{equation}
which quantifies the amount of turbulence over a characteristic length scale $D$ (in this case taken to be the size of the convection zone), where $v_\mathrm{rms}^\mathrm{CZ}$ is the rms velocity averaged over the convection zone; 
and the \textit{Prandtl number}
\begin{equation}
    \mathrm{Pr} = \frac{\nu}{\kappa},
\end{equation}
which indicates whether momentum ($\nu$) or heat ($\kappa$) dissipates more efficiently within the given medium.  
These two characteristic dimensionless quantities help assess our simulations against real stellar flows. 
Numerical methods are unfortunately unable to perfectly emulate stellar diffusive properties due to stability restrictions, with considerably higher values needed for both the kinematic viscosity $\nu$ and the thermal diffusivity $\kappa$. 
Using the ZAMS model as an example, stellar values for the thermal diffusivity within the simulation domain range from $\kappa\approx 10^8 \, \si{cm^2 s^{-1}}$ to $\approx 10^{13} \, \si{cm^2 s^{-1}}$, while the stellar kinematic viscosity spans from $\nu_\star \approx 10^3 \, \si{cm^2 s^{-1}}$  to $\approx 10^{6} \, \si{cm^2 s^{-1}}$. The enhanced strength of the diffusive processes in the simulations would however dampen waves too much to allow any meaningful analysis and comparisons with observations, so we attempt to circumvent this issue by amplifying the stellar luminosity (and therefore convective velocities) by a factor of $\sim 10^{4}$ (corresponding to a boost of $\sim 10^{1.33}$ in $v_\mathrm{rms}^\mathrm{CZ}$ as suggested by eg. \cite{PorterWoodward2000,Vialletetal2013}). 

Despite the enhanced luminosity, our simulation flows still fall short of matching the stellar $\mathrm{Re}$
, with stellar value being $\mathrm{Re}_\star \sim 10^{12}$ 
and our models reaching $\mathrm{Re} \sim 510-735$ 
, indicating simulations cannot replicate the amount of turbulence found within stellar interiors. 
The restrictions on $\nu$ due to numerical stability also imply the simulation flows have a much larger $\mathrm{Pr}$ than real stars, with MESA models possessing $\mathrm{Pr}_\star \sim 10^{-6}$ in the core and $\sim 10^{-9}$ in the envelope, while the respective values for our simulations are $\mathrm{Pr} \sim 5-15$ in the CZ and $\mathrm{Pr} \sim 0.02-0.4$ in the envelope; the only way to improve this would have been to substantially increase $\kappa$ further, which would unfortunately result in excessive wave damping/attenuation as explained above. 
Regardless, this still represents a mild improvement compared to the values of $\mathrm{Re}$ and $\mathrm{Pr}$ reached in \cite{Edelmannetal2019}, with this work achieving Reynolds numbers larger by roughly one order of magnitude and Prandtl numbers that are smaller by a comparable amount. 

Given that the aim of our work is to compare simulations of different stellar ages as well as to highlight and explain any resulting differences,  we have kept similar $\nu$ and $\kappa$ profiles across all ages. We have, therefore, not investigated the role of $\nu$ and $\kappa$ on the results. Instead, we wish to highlight the dynamical differences as a function of stellar age and focus on the non-damping causes driving them. Any such non-damping effects are likely to also be relevant in real stars, where damping mechanisms are far less important than in these simulations.



\section{Results} \label{sec:results}
\begin{figure}[t] 
	\includegraphics[width=1.\columnwidth]{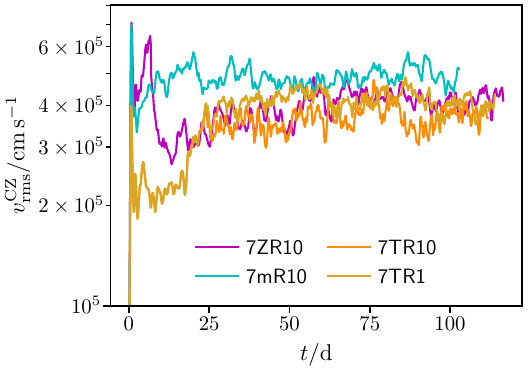}
	\caption{Temporal evolution of the $v_\mathrm{rms}$ averaged over the entirety of the convective zone for each model.}
	\label{fig:vrms_avg_CZ_tevo}
\end{figure}

\begin{figure*}[ht] 
	\centering
	\subfloat{%
	    \includegraphics[width=\columnwidth]{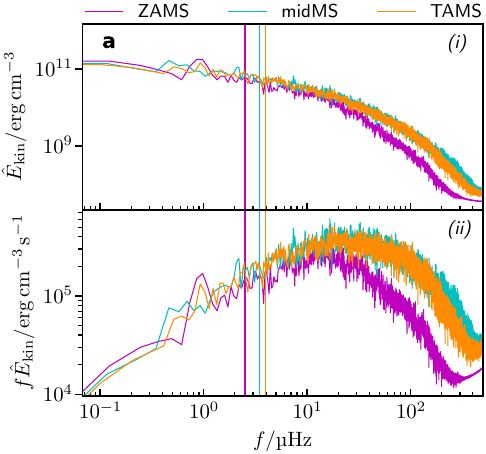}
	    \label{fig:generation-spectra-inside-current-CZ} }
	\hspace*{3em}
	\subfloat{%
	    \includegraphics[width=\columnwidth]{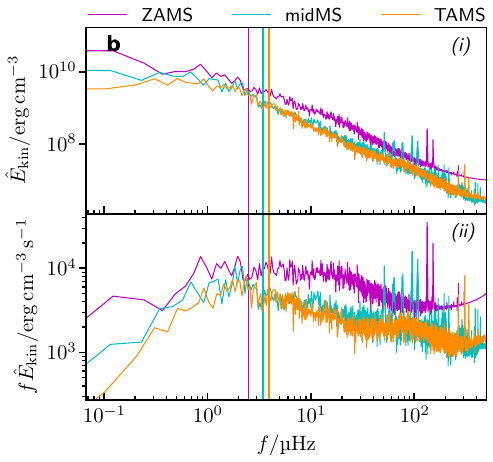} 
	    \label{fig:generation-spectra-outside-current-CZ} }
	\caption{Kinetic energy frequency spectra for all three stellar ages just below (\textbf{\textit{a}}) and above (\textbf{\textit{b}}) the convective boundary. For each radial location, the Fourier transform of the kinetic energy (from Eq. \ref{eq:KE-Fourier}) presented in the top panel \textit{(i)} is multiplied by the frequency $f$ in the bottom panel \textit{(ii)} to compensate for the integration over $\mathrm{d} \log f$. The colour-coded vertical lines represent estimates for the turnover frequencies of the models.}
	\label{fig:generation-spectra-in-out-current-CZ}
\end{figure*}

We show the temporal evolution of the $v_\mathrm{rms}$ averaged within the entirety of the convective zone for each of the models presented in Fig.~\ref{fig:vrms_avg_CZ_tevo}. All models feature an initial rapid increase in CZ velocities, which signals the onset of convection; these velocities are then seen to settle down into a steady state. All of the following analysis is done in this steady-state region. The steady-state convective velocities are very similar across all models, with only a factor of $\sim1.3$ between the highest (7mR10) and the lowest (7TR10) temporally averaged convective velocities; this is by design, as the luminosity boosting was gently tweaked to ensure convective velocities would be similar for all stellar ages to be able to effectively compare IGW propagation as a function of stellar age. We note that MLT velocities from MESA show younger stars with slightly lower velocities than their older counterparts (by about 1.2 from TAMS to ZAMS).

\subsection{IGW generation} \label{sec:results-generation}

As mentioned in Section~\ref{sec:intro}, theoretical studies on IGW propagation within stars assume that their generation spectra are driven exclusively by either overshooting plume incursions \citep{Townsend1966, MontalbanSchatzman2000, Pinconetal2016} 
or convective eddies \citep{Lighthill1952, GoldreichKumar1990, Kumaretal1999, LecoanetQuataert2013}. More recent analysis by \cite{LeSauxetal2023} notes that the dominance of plumes or eddies may be related to the convective forcing. Of course, it is highly likely that both plumes and convective eddies play a part in generating IGWs.  Given that the spectra of IGWs throughout the star are sensitive to their generation mechanism, it is important to understand the relevance of both of these mechanisms within our simulations. For this reason, we analyse frequency spectra for the kinetic energy 

\begin{equation} \label{eq:KE-Fourier}
    \hat{E}_\mathrm{kin} = \frac{1}{2} \overline{\rho} \left(\hat{v}_r^2 + \hat{v}_\theta^2 + \hat{v}_\phi^2 \right), 
\end{equation}
where $\hat{E}_\mathrm{kin}$ represents the Fourier transform of the real quantity $E_\mathrm{kin}$. 


Figure~\ref{fig:generation-spectra-in-out-current-CZ} shows the frequency spectra for the kinetic energy $\hat{E}_\mathrm{kin}$ taken $0.07 H_p$ below (Fig.~\ref{fig:generation-spectra-inside-current-CZ}) and $0.07 H_p$ above (Fig.~\ref{fig:generation-spectra-outside-current-CZ}) the convective boundary\footnote{the pressure scale height is defined as $H_p=-\partial r/\partial \ln P$} (the latter radial location is chosen to match the value used for the generation spectra analysis by \cite{Edelmannetal2019}, which has the added advantage of being located within the Brunt-Vaisala frequency spike in midMS and TAMS, allowing us to sample all the waves generated) for all three stellar ages in panel \textit{(i)}. In panel \textit{(ii)} the same kinetic energy spectrum has been multiplied by the frequency $f$ to account for the integration over $d\log f$; this helps to emphasise the regions where most kinetic energy is contained. In both panels we also plot colour-coded vertical lines corresponding to the estimates of the turnover frequencies of the models according to

\begin{equation}
    f_\mathrm{TO} = \frac{v_\mathrm{rms}^\mathrm{CZ}}{\pi r_\mathrm{CZ}} ,
\end{equation}
where the expression assumes that the largest eddy spans the entire radial extent of the convective zone $r_\mathrm{CZ}$ and turns at $v_\mathrm{rms}^\mathrm{CZ}$; for convenience we also note here the actual values of the turnover frequencies used, which are: $f_\mathrm{TO}^\mathrm{Z}=2.51\,\si{\micro Hz}$, $f_\mathrm{TO}^\mathrm{m}=3.45\,\si{\micro Hz}$ and $f_\mathrm{TO}^\mathrm{T}=3.93\,\si{\micro Hz}$.
In Fig.~\ref{fig:generation-spectra-inside-current-CZ} the peaks of the spectra in panel \textit{(ii)} are located at frequencies roughly one order of magnitude higher than the estimated turnover frequencies; furthermore, the spectra do not appear dominated by one single frequency, with most of the kinetic energy of the motions being stored in frequencies within the range $10^1 \lesssim f \lesssim 10^2 \, \si{\micro Hz}$. This is in contrast with the theoretical idea of a generation spectrum that is mostly influenced by convective eddies, where most of the energy is concentrated at the convective turnover frequency with steep drops away from it. While the lower end of the spectrum within the radiative zone could be affected by viscous and thermal diffusivities, the fact that we don't see a prominent convective frequency within the convection zone indicates that this concept may not be relevant in this geometry.

The spectra shown in Fig.~\ref{fig:generation-spectra-outside-current-CZ}, as mentioned previously, are taken just above ($0.07 H_p$) the CB rather than below it; in the midMS and TAMS models this approximately coincides with the maximum value reached by $N$ in their BVF spikes. Here the spectra, particularly once they are multiplied by $f$ in panel \textit{(ii)}, appear much flatter and it is impossible to pinpoint a dominant frequency (particularly for the ZAMS model) or the location of the turnover frequencies. 

While both of these sets of spectra represent the motions generated at the CB, rather than specifically waves, these motions are responsible for driving the waves and the properties of the two are therefore closely linked; given the spectra of the motions on either side of the CB are not dominated by the turnover frequency, it is safe to assume that the IGW energy will likewise not be concentrated at $f_\mathrm{TO}$.

\begin{figure*}[th] 
	\includegraphics[width=1.\textwidth]{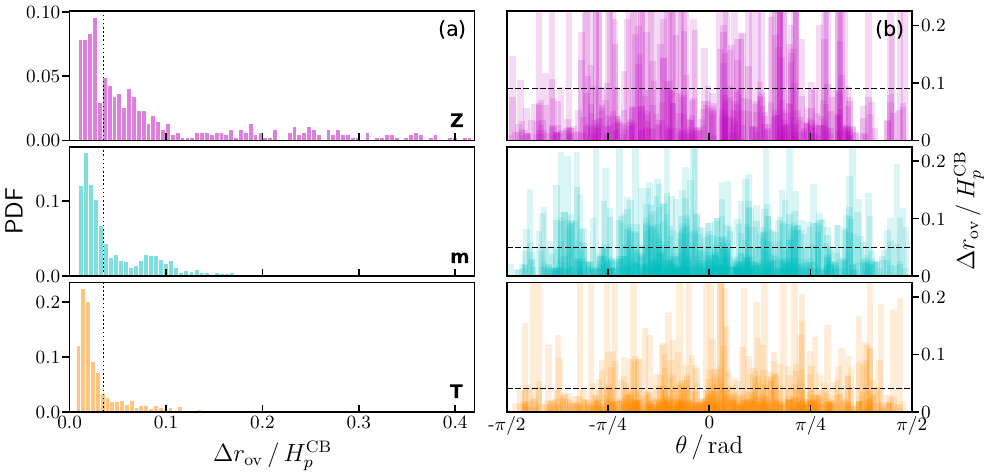}
	\caption{Normalised PDFs of the plume overshooting length in units of the pressure scale height (panel \textit{(a)}) and latitudinal distribution of plumes (panel \textit{(b)}) for each stellar age (Z: ZAMS, purple; m: midMS, cyan; T: TAMS, orange)}
	\label{fig:models-plume-depths-hist-and-theta}
\end{figure*}

\subsection{Overshooting plumes} \label{sec:results-plumes}
Convective plumes overshooting into the radiative zone play a significant role in the dynamical evolution of stars. Apart from generating IGWs at the convective boundary, as previously mentioned, they also lead to the formation of a layer between the convective and radiative zones where turbulent motions intermittently mix with the radiative flow. The properties of this layer can influence chemical mixing, the generation of the magnetic field and the evolutionary track in the Hertzsprung–Russell diagram of a star, therefore playing a role in its evolution \citep{MarikPetrovay2002,Marquesetal2006,Tianetal2009}. 

While the physics of this overshooting layer lie beyond the scope of this paper, we aim to understand how plume overshooting varies with age across our models. For this purpose, we carry out an analysis on their incursion into the radiation zone; we scan every output file (which are separated by an interval of $1000\, \si{s}$) in every ($\phi, \theta$) direction to look for possible plumes overshooting beyond the convective boundary. This is accomplished by monitoring the sign of the radial velocity $v_r$ or, equivalently, of the vertical kinetic energy flux $E_K = \frac{1}{2} v_r \rho \boldsymbol{v}^2$ \citep{Hurlburtetal1986,Hurlburtetal1994}, whose sign changes when plumes are decelerated. 
In particular, we define the overshooting length $\Delta r_\mathrm{ov}$ of a plume at a given $\phi$, $\theta$ combination as the distance from the CB at which $v_r$ turns from positive to negative. 
The results are subsequently refined to remove false positives by filtering all the obtained overshooting lengths by a minimum depth (ie. the radial extent of a few grid cells, meaning any smaller depths are ignored) and by finding structures which are coherent both spatially and temporally among the filtered data. Each coherent structure is labelled as a plume, and its maximum incursion depth and velocity are recorded. 

A similar analysis has been carried out by \cite{Prattetal2016,Prattetal2017}, who employed 2D simulations of spherical shells of various thicknesses to study overshooting over several tens of convective turnover times or more. Apart from the differences in the dimensional and geometrical setup, the main difference is that the analysis by \cite{Prattetal2016,Prattetal2017} is focused on a proto-type of a young Sun, where plumes are launched inward, along an increasing density gradient, from a sub-surface convection zone. Of course the opposite is true in our models, where plumes are launched outwards from a convective core.


Figure~\ref{fig:models-plume-depths-hist-and-theta} shows the probability density functions (PDFs) of the plume overshooting length in units of the pressure scale height at the convective boundary $H_p^\mathrm{CB}$ (panel \textit{(a)}) for each stellar age (Z: ZAMS, purple; m: midMS, cyan; T: TAMS, orange), and an illustration of the overshoot of plumes in our simulations as a function of the latitude $\theta$ (panel \textit{(b)}). For reference, the pressure scale heights at the CB for the three models are $H_{p,\mathrm{Z}}^{\mathrm{CB}}= 0.095 R_\star$, $H_{p,\mathrm{m}}^{\mathrm{CB}}= 0.079 R_\star$ and $H_{p,\mathrm{T}}^{\mathrm{CB}}= 0.052 R_\star$.

The PDFs presented in panel \textit{(a)}, which have been normalised such that the heights of their bars sum up to 1, show that for all stellar ages a large fraction of plumes only reach an approximate maximum overshooting length of $\Delta r_\mathrm{ov}/H_p^\mathrm{CB} = 0.035$ (vertical dotted lines). There also appears to be a second overshoot layer, directly above the first, which plumes reach more infrequently. This deeper overshoot layer, which is also observed in the 2D simulations by \cite{Prattetal2017}, appears fairly accessible in the ZAMS case but becomes harder to reach with increasing stellar age; few plumes are able to overshoot beyond $\Delta r_\mathrm{ov}/H_p^\mathrm{CB} = 0.035$ in the TAMS case. This can be quantified by calculating the percentage of plumes overshooting beyond $\Delta r_\mathrm{ov}/H_p^\mathrm{CB} = 0.035$ for the three ages: while 67$\%$ of plumes overshoot beyond this threshold for ZAMS, this figure drops to $43\%$ and $31\%$ for midMS and TAMS, respectively. While the ZAMS value is probably slightly inflated by some potential false positives found beyond $\Delta r_\mathrm{ov}/H_p^\mathrm{CB} \sim 0.2$, the downward trend with stellar age is clear.
The reason for this decline in the number of plumes overshooting beyond this threshold, given that the three models possess comparable generation spectra (Fig.~\ref{fig:generation-spectra-inside-current-CZ}), can be safely attributed to the emergence of the spike in the Brunt--V\"{a}is\"{a}l\"{a} frequency in older stars, caused by a steep compositional gradient. 



Panel \textit{(b)} of Fig.~\ref{fig:models-plume-depths-hist-and-theta} shows the latitudinal distribution of overshooting plumes for each model as a function of the normalised overshooting length. Here each vertical bar represents a plume (the plume size is not taken into consideration, therefore all bars have equal width), with overlapping plumes resulting in darker regions. There appears to be no preferential latitude for plume overshoot, although fewer plumes are observed near the stellar poles. The mean overshooting length is also indicated for each model by means of a horizontal dashed line; this appears to decrease with stellar age, with a significant drop from ZAMS to midMS (from $\Delta r/H_p^\mathrm{CB} \approx 0.09$ to $\approx 0.05$), and a minor one from midMS to TAMS (from $\approx 0.05$ to $\approx 0.04$). These results are comparable with the results from the 2D simulations by \cite{Prattetal2017}, as well as the 3D simulations of a $2M_\odot$ A-star by \cite{Browningetal2004}; in the latter, the authors find overshooting depths of roughly $0.2H_p$, and notice the overshooting length shrinking with increasing Reynolds number. This would potentially reconcile the results even more, as they consider Reynolds numbers that are $2.5-50$ times smaller than the one considered in our ZAMS model. 
While overshooting lengths across all models could potentially be inflated by the boosted stellar luminosities used, the observed trend is consistent with the percentage of plumes overshooting to the second overshooting layer mentioned above. 

\subsection{Which is dominant: plumes or eddies?} \label{sec:results-dominance}
In order to understand the relative importance of overshooting plumes and convective eddies in our generation spectra, we carry out a direct comparison between the spectra shown in Fig.~\ref{fig:generation-spectra-in-out-current-CZ} and the analytical prescriptions for both formalisms. 
To begin with, we consider two examples of eddy-dominated generation spectra from \cite{Kumaretal1999} (labelled K) and \cite{LecoanetQuataert2013} (labelled LD for ``Lecoanet Discontinuous", to indicate it is the profile obtained with a discontinuous stratification profile at the convective boundary) according to which the wave energy spectrum is described by the following power laws:
\begin{equation} \label{eq:powerlaw-kumar} 
    E_\mathrm{kin}^\mathrm{K} \propto f^{-4.34}, 
\end{equation}

\begin{equation}\label{eq:powerlaw-lecoanet}
    E_\mathrm{kin}^\mathrm{LD} \propto f^{-6.5}. 
\end{equation}
We also consider a third power law spectrum, obtained by \cite{Rogersetal2013} (labelled R) from 2D simulations of a $3M_\odot$ star. Their spectra are actually best described by two distinct power laws; for simplicity we pick the power law exponent stretching the largest portion of the spectrum, encompassing from low frequencies up to $f\sim 100\,\si{\micro Hz}$. Since \cite{Rogersetal2013} do not analyse the same rotation rates used here (the models in both Fig~\ref{fig:generation-spectra-inside-current-CZ} and Fig.~\ref{fig:generation-spectra-outside-current-CZ} have $\Omega=10^{-5}\,\si{rad \, s^{-1}}$), we take an average of their results obtained with the two closest rotation rates, resulting in
\begin{equation} \label{eq:powerlaw-rogers}
    E_\mathrm{kin}^\mathrm{R} \propto f^{-0.85} .
\end{equation}

With regards to IGWs generated by plume overshoot, we adopt the formalisms developed by \cite{MontalbanSchatzman2000} and \cite{Pinconetal2016}, which are based on the plume models by \cite{RieutordZahn1995} and \cite{Zahn1991}. These studies 
are only valid at the high P\'{e}clet numbers (ie. $\mathrm{Pe}>> 1$), with the P\'{e}clet number being defined as 
\begin{equation*}
    \mathrm{Pe} = \frac{uL}{\kappa},  
\end{equation*}
where $u$ and $L$ are the characteristic velocity and length scale of the convective motions. 
At the radii of the spectra shown in Fig.~\ref{fig:generation-spectra-in-out-current-CZ} all our models show $\mathrm{Pe}$ values of order $\sim 10^4$, 
allowing us to apply their formalism with confidence. The plume energy spectrum depends on the plume timescale $\tau_\mathrm{pl}$, which is approximated as 

\begin{equation}
    \tau_\mathrm{pl} \sim \frac{\Delta r_\mathrm{ov}}{v_\mathrm{pl}},
\end{equation}
where $v_\mathrm{pl}$ is the velocity of a plume and $\Delta r_\mathrm{ov}$ its overshooting length. Its reciprocal gives the plume frequency

\begin{equation} \label{eq:plume-freq}
    f_\mathrm{pl} \sim \frac{v_\mathrm{pl}}{\Delta r_\mathrm{ov}};
\end{equation}
the energy spectrum takes an exponential form in both \cite{MontalbanSchatzman2000} and \cite{Pinconetal2016} cases, with the general form of

\begin{equation} \label{eq:plume}
    E_\mathrm{kin}\left(f\right) \propto e^{-\left(f/f_\mathrm{pl}\right)^2}.
\end{equation}

\begin{figure}[t] 
	\includegraphics[width=1.\columnwidth]{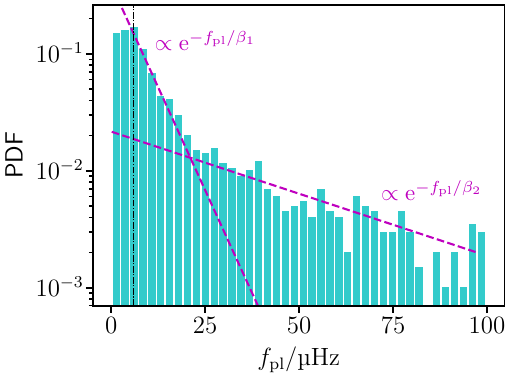}
	\caption{Normalised PDF of the plume frequency $f_\mathrm{pl}$ for the midMS model. The vertical, black dot-dashed line represents the peak frequency, located at $f_\mathrm{pl} \approx 6\si{\micro Hz}$; the magenta straight lines, possessing slopes of $-1/\beta_1$ and $-1/\beta_2$, represent exponential fits to the data.}
	\label{fig:plume-frequency-PDF-ZAMS}
\end{figure}

While this simplified expression yields the kinetic energy generation spectrum for a specific plume frequency, it does not provide any information about the frequency distribution of such plumes; each of our simulations of course possesses a unique $f_\mathrm{pl}$ distribution, obtained from the plume analysis described in Section~\ref{sec:results-plumes}, with the normalized plume frequency PDF for the midMS model shown in Fig.~\ref{fig:plume-frequency-PDF-ZAMS}. The distribution peaks at $\approx 6 \, \si{\micro Hz}$ (black, dot-dashed vertical line) and appears well described by two exponential fits (dashed magenta lines) having gradients of $-1/\beta_1$ and $-1/\beta_2$, with $\beta_1 = 6.2\, \si{\micro Hz}$ and $\beta_2 = 40.9 \, \si{\micro Hz}$, with a cross-over point at roughly $f_\mathrm{pl} \approx 20 \, \si{\micro Hz}$. The $f_\mathrm{pl}$ distributions of the other models are similarly well described by two exponential fits. 
We therefore combine the double exponential nature of the $f_\mathrm{pl}$ distributions from our models with the original plume prescriptions by \cite{MontalbanSchatzman2000} (labelled M) and \cite{Pinconetal2016} (labelled P) to ensure the appropriate plume frequency contributions are considered. This is achieved through an extra $e^{-f_\mathrm{pl}/\beta_n}$ factor (with $n=1,2$);
the adapted plume prescriptions become

\begin{figure*}[!t]
    \centering
    \includegraphics[width=\textwidth]{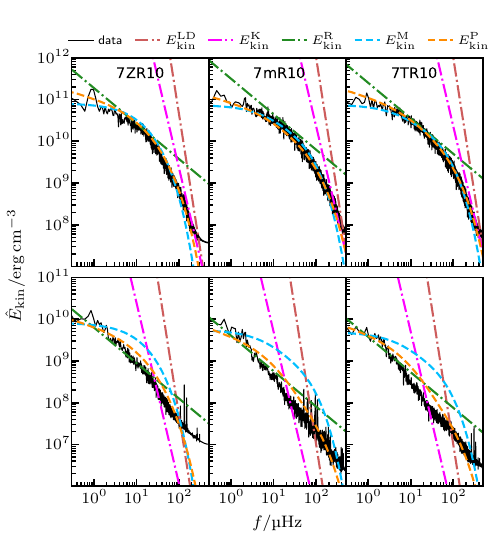}
    \caption{Comparison between simulation spectra (black, full lines) taken $0.07H_P$ below and above the CB (top and bottom panels, respectively), with power law generation spectrum profiles generated by turbulent eddies (dot-dashed, maroon and magenta), a power law generation spectrum from a 2D hydrodynamical simulation (green dot-dashed), and two exponential generation spectrum profiles caused by overshooting convective plumes (dashed lines, blue and orange).}
    \label{fig:generation-spectra-both-inside-outside-current-CZ-mechanism-comparison}
\end{figure*}



\begin{equation} \label{eq:plume-integral-montalban} 
     E^\mathrm{M}_\mathrm{kin}\left(f\right) \propto \sum_{n} \int_{0}^{\infty} e^{-f_\mathrm{pl}/\beta_n} \, e^{-\left(f/f_\mathrm{pl}\right)^2} \mathrm{d}f_\mathrm{pl} ,
\end{equation}

\begin{equation} \label{eq:plume-integral-pincon} 
     E^\mathrm{P}_\mathrm{kin}\left(f\right) \propto \sum_n \int_{0}^{\infty} e^{-f_\mathrm{pl}/\beta_n} \, f_\mathrm{pl}^{-1}e^{-\left(f/4f_\mathrm{pl}\right)^2} \mathrm{d}f_\mathrm{pl} ,
\end{equation}
where $-1/\beta_n$ is the slope of each exponential fit to the unique $f_\mathrm{pl}$ PDF of each model, as shown in Fig.~\ref{fig:plume-frequency-PDF-ZAMS} by the dashed magenta lines. 

Figure~\ref{fig:generation-spectra-both-inside-outside-current-CZ-mechanism-comparison} showcases the results of the comparison between our simulation spectra (black, full lines) and Eqs.~\ref{eq:powerlaw-kumar}--\ref{eq:powerlaw-rogers}, \ref{eq:plume-integral-montalban}--\ref{eq:plume-integral-pincon} for all three stellar ages, both just below (top panel) and above (bottom panel) the convective boundary.
Below the convective boundary (top panel), the adapted plume prescriptions appear to describe the generation spectra exceedingly well throughout most of the frequency range and for all three stellar models. 
Out of the two plume generation spectra analysed, the \cite{Pinconetal2016} prescription $E_\mathrm{kin}^\mathrm{P}$ (orange dashed line) appears the best fit to the simulation data, while $E_\mathrm{kin}^\mathrm{M}$ (blue, dashed) appears to produce a profile whose gradient changes too sharply around $f\approx 10\si{\micro Hz}$ compared to the simulation spectra, resulting in either an excessive drop in amplitude at high frequencies, or a profile that is too flat in the low $f$ range. This result is in line with the findings by \cite{Edelmannetal2019}, who also found their generation spectra to be well described by plume-driven motions.

It is worth noting that all three spectra are also sufficiently well described by a broken power law with the best fits among the power laws used being $E_\mathrm{kin}^\mathrm{R}$ (green, dot-dashed) for low $f$ and $E_\mathrm{kin}^\mathrm{K}$ (magenta, dot-dashed) for high frequencies (although both of these power laws look somewhat too steep for our data). This is well aligned with the numerical results of \cite{Rogersetal2013} which also found a broken power law with exponents of $\alpha\sim 1$ for low and intermediate frequencies and $\alpha \sim4$ (therefore very similar to the power law exponent in $E_\mathrm{kin}^\mathrm{K}$) for high $f$ for stellar rotation rates close to that of the three models shown here. However, it is clear that this broken power law fit does not represent as accurate a description as the plume prescriptions, particularly as it appears to substantially overestimate the spectrum at low frequencies.


At the location just above the CB (bottom panel), the situation is somewhat different. All three spectra have been affected in the mid-- and high-frequency ranges, and their profiles now appear better described by a single power law with an index similar to that of $E_\mathrm{kin}^\mathrm{R}$. While this is true for all models, the ZAMS spectrum still appears adequately well described by a \cite{Pinconetal2016} plume-driven profile; however, the agreement between the spectra and their respective $E^\mathrm{P}_\mathrm{kin}$ fits deteriorates with stellar age. This may be expected given those theoretical models do not account for BVF spikes as the star ages.

\begin{figure*}[t]
    \centering
    \includegraphics[width=0.9\textwidth]{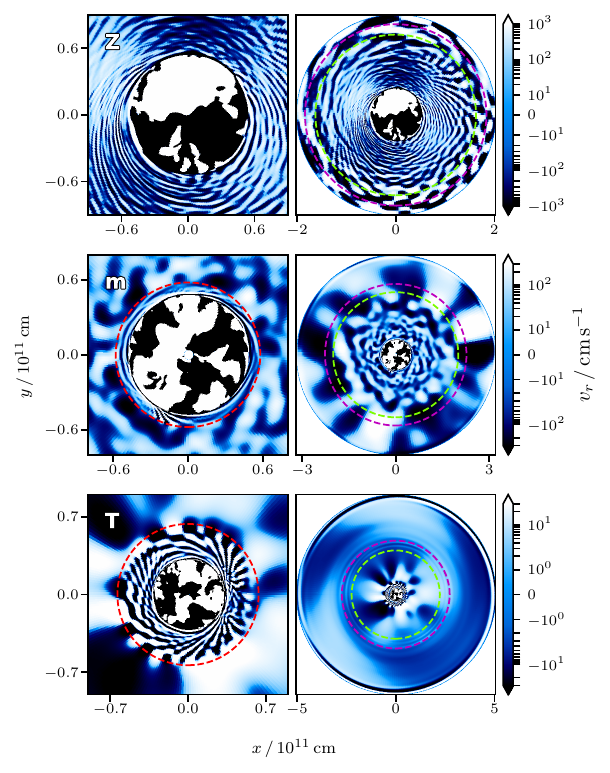}
        \caption{Equatorial slice comparison for the radial velocity $v_r$ across the three stellar ages: ZAMS (Z, top row), midMS (m, middle row) and TAMS (T, bottom row). The slices on the right represent the view of the whole domain, while the slices on the left show a zoomed in view of the convection zones and their immediate surroundings. The extent of the BVF spikes in the midMS and TAMS models is also indicated on the left panels by means of red dashed circles. The green and magenta dashed circles on the right panels indicate the radial locations at which waves with $15\, \si{\micro Hz}$ (green, inner circle) and $10\, \si{\micro Hz}$ (magenta, outer) frequencies (with $m=1$) become evanescent according to the 2D linear analysis by \cite{Ratnasingametal2020}.}
    \label{fig:CZ-RZ-comparison}
\end{figure*}

\begin{figure*}[t] 
	\includegraphics[width=1\textwidth]{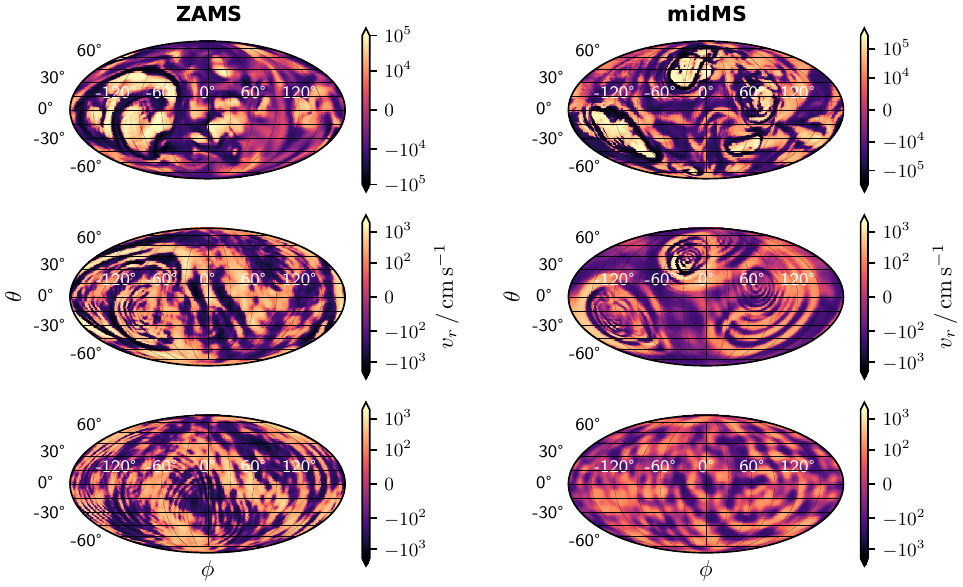}
	\centering
	\caption{Mollweide projections at constant radius for the ZAMS (left) and midMS (right) models, taken at three radial locations: at the CB (top row), and at $0.02R_\star$ (middle row) and $0.15R_\star$ (bottom row) above it. We do not show TAMS as there is very little structure outside the BVF spike.}
	\label{fig:mollweide_midMS_ZAMS}
\end{figure*}

\subsection{IGW propagation} \label{sec:results-propagation}
We now move on to consider the propagation of waves throughout the entirety of the radiation zone. Figure~\ref{fig:CZ-RZ-comparison} shows equatorial slices illustrating the propagation of IGWs through the RZ across the three stellar ages: ZAMS (labelled Z, top row), midMS (m, middle row) and TAMS (T, bottom row). While in the ZAMS model the generated IGWs are seen to freely propagate throughout the whole radiation zone, in the midMS and TAMS models the propagation of waves is significantly affected beyond their BVF spikes (the extent of which is marked by red dashed circles), 
leaving the rest of their RZs more sparsely populated by IGWs. 
This is further highlighted by the velocity amplitudes of the slices, which clearly decrease with age (see associated colorbar). 

The depletion of IGWs in the top part of the RZ of older stars is due to a combination of three separate effects. The one we believe to be the most important, both in these simulations and in real stars, is IGWs becoming evanescent as they propagate towards the stellar surface. As explained by \cite{Ratnasingametal2020} in their 2D cylindrical model, the 2D linearised anelastic equations can be reduced to a single second order differential equation describing the evolution of IGWs

\begin{equation}\label{eq:turning-point-full}\begin{multlined}
    \frac{\partial^2 a}{\partial r^2}  + \left( \frac{N^2}{\omega^2} -1 \right)k_h^2 a \\
         + \frac{1}{2} \left[\frac{\partial h_\rho}{\partial r} - \frac{h_\rho^2}{2} + \frac{h_\rho}{r} \right] a  + \frac{1}{4r^2} a = 0 \nonumber,
\end{multlined}\end{equation}
where $a=v_r \overline{\rho}^{1/2}r^{3/2}$. 
IGWs maintain their wave-like properties as long as the ratio between the oscillatory term (OT, second term lhs of equation) and the density term (DT, third term lhs of equation) remains above unity, ie.

\begin{equation} \label{eq:turning-point}
    \frac{OT}{DT} = \frac{\left( \frac{N^2}{\omega^2} -1 \right) \frac{m^2}{r^2} }{\frac{1}{2} \left[\frac{\partial h_\rho}{\partial r} - \frac{h_\rho^2}{2} + \frac{h_\rho}{r} \right]} > 1.
\end{equation}
Within the radiation zone the value of this ratio decreases with radius, potentially reaching values below one depending on the frequency chosen and the value of $m$. The radius at which the ratio reaches unity is called the turning point, and IGWs lose their wave-like behaviour beyond it. \cite{Ratnasingametal2020} also observed the location of the turning point receding to smaller $r/R_\star$ values with increasing age for waves of the same frequency, with IGWs in older stars therefore becoming evanescent earlier in their propagation, hence contributing to the scarcity of IGWs observed in our midMS and TAMS models. The reason older stars are more susceptible to this phenomenon is the overall decrease in the Brunt--V\"{a}is\"{a}l\"{a} frequency as a function of stellar age, as seen in Fig.~\ref{fig:sims_N2}, which results in a decreased oscillatory term OT for all frequencies. To put the magnitude of this effect into perspective, a wave of frequency $f=15 \, \si{\micro Hz}$ ($m=1$) loses its wave-like behaviour past $r/R_\star\approx 0.72$ for ZAMS, but its turning point recedes significantly to $r/R_\star \approx 0.56$ and $r/R_\star \approx 0.38$ for midMS and TAMS, respectively. 
This effect, which has not been considered before when analyzing the propagation of IGWs in stars and their impact on stellar dynamics and observational signatures, is believed to also be responsible for the change in wave structure seen across the RZ in Fig.~\ref{fig:CZ-RZ-comparison} as a progressively increasing fraction of IGWs become evanescent as a function of radius. This is supported by the magenta and green dashed circles on the right panels of Fig.~\ref{fig:CZ-RZ-comparison}, which represent the locations at which $m=1$ waves with frequencies of $15\,\si{\micro Hz}$ (green, inner circle) and $10 \,\si{\micro Hz}$ (magenta, outer)\footnote{These frequencies were chosen as an example as they are able to freely propagate in the entirety of the stellar radiative zone in all three models, and are not excessively damped by diffusion.} become evanescent in each model as predicted by Eq.~\ref{eq:turning-point}, roughly coinciding with the location where the wave structure changes. While we see this effect in our simulations, it is likely to be much more relevant in actual stars, where viscous and thermal damping are far less relevant.  Moreover being a linear effect, it is unaffected by most of the simulation assumptions and shortcomings.

On top of IGW propagation in older stars being more restricted due to turning point limitations, a substantial portion of generated IGW frequencies in the midMS and TAMS models become trapped within the respective BVF spikes; this is caused by the drop in $N$ at the end of the BVF spike, which causes high-frequency IGWs to no longer be permitted to propagate as $\omega < N$ would be violated. This phenomenon is visualised in the left panels of Fig.~\ref{fig:CZ-RZ-comparison} for midMS and TAMS, where it is possible to see IGWs trapped within the BVF spike (the extent of which is marked by the dashed red circles). There we see the angle of the phase lines change dramatically indicating reflection. As shown in the  Brunt--V\"{a}is\"{a}l\"{a} frequency profile from Fig.~\ref{fig:sims_N2}, for TAMS this jump in $N^2$ is of a factor of $\approx 30$, meaning frequencies in the $70 \lesssim f \lesssim 400 \, \si{\micro Hz}$ range are trapped within the BVF spike. 

Lastly, the enhanced radiative damping within the BVF spike due to the large value of $N$ within the spike itself 
also contributes to the scarcity of IGWs in the top part of the radiative zones in midMS and TAMS. Linear theory suggests velocity amplitudes of waves propagating through the radiation zone will evolve according to several effects, such as density stratification, geometrical effects and radiative damping. \cite{Ratnasingametal2019}, building on the work by \cite{Press1981} and \cite{Kumaretal1999}, provide the following expression for the linear wave velocity amplitude
\begin{equation} \label{eq:radiative-diff}
    v_r \propto \left( \frac{r_0}{r} \right)^{3/2} \sqrt{\frac{\rho_0}{\rho}} \left(  \frac{N^2-\omega^2}{N_0^2-\omega^2} \right)^{1/4} e^{-\tau/2} ,
\end{equation}
with $\omega=2\pi f$, where $r_0$ is the radius at which waves begin propagating, and $N_0$ and $\rho_0$ are the Brunt--V\"{a}is\"{a}l\"{a} frequency and density at $r_0$; $\tau$ is the damping opacity, whose full expression is 
\begin{equation} \label{eq:radiative-damping-length} \begin{multlined}
    \tau = \int_{r_0}^r \frac{\gamma_\mathrm{rad}}{\lvert v_\mathrm{g} \rvert} \, \mathrm{d}r, \nonumber \\ 
         = \int_{r_0}^r \frac{\kappa \left[ l \left( l+1 \right) \right]^{3/2}}{r^3 \omega^4} N^3 \sqrt{1-\frac{\omega^2}{N^2}} \, \mathrm{d}r,
\end{multlined}\end{equation}
where $\gamma_\mathrm{rad}$ is the radiative damping rate and $v_g$ the wave group velocity. 
The amount of radiative damping on waves is therefore very sensitive to the values of $N$, implying that a specific frequency $\omega$ of a given $l$ value would be significantly more damped within the BVF spike of an older star than in the corresponding radial location of a younger one due to the higher value of $N$. 
To quantify this increased radiative damping within the BVF spike of older stars, we use Eq.~\ref{eq:radiative-damping-length} to calculate the damping factor $\exp(-\tau/2)$ for a $f=10\,\si{\micro Hz}$ wave with $l=2$ at $r=0.5R_\star$ for the ZAMS and TAMS models. We find that at the given location, the TAMS wave is subject to a damping that is $\sim 5$ times stronger than its ZAMS counterpart. Radiative damping for a fixed $l$ value is at its strongest in low frequency waves ($\omega/N \ll 1$) since $\gamma_\mathrm{rad}/\lvert v_g \rvert \propto \omega^{-4}$, meaning these waves would be particularly affected.With \cite{Talonetal2002} treating viscosity and thermal diffusion as additive when considering wave damping, it is possible viscosity dampens IGWs in our simulations at least as much as $\kappa$; in any case, comparing radial velocity amplitudes from our simulations to those obtained using stellar $\kappa$ values, we find waves with $f\gtrsim 5\si{\micro Hz}$ and $m \lesssim 5$ are consistent with MESA. However, it is important to note that while changes in radiative damping as a function of age must be considered in our simulations due to the excessive thermal diffusivities used for numerical reasons, we do not expect this phenomenon to be significant in real stars. As real stars feature much smaller values for $\kappa$, the changes in radiative damping across different stellar ages is negligible for most wavenumber and frequency permutations unless considered near the stellar surface. Our focus is, therefore, centred on the turning point and the IGW trapping mechanisms, both non-damping in nature, which also play an important role in IGW propagation in real stars.

\begin{figure*}[t] 
	\includegraphics[width=\textwidth]{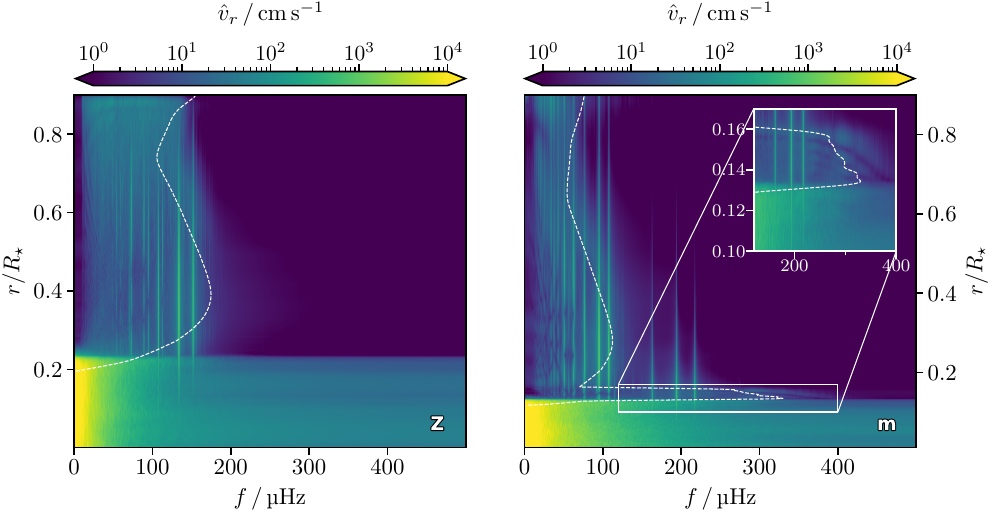}
    \caption{Frequency spectrum heat maps for all radii of the radial velocity $v_r$, integrated over all $l$, calculated at the stellar equator for models 7ZR10 (Z, left) and 7mR10 (m, right). The dashed white lines represents the Brunt--V\"{a}is\"{a}l\"{a} frequency $N/2\pi$, which approximately constrains the signal as expected for IGWs. The inset axes in the 7mR10 plot show a zoom on the BVF spike} 
    \label{fig:ZAMS_midMS_spectral_map}
\end{figure*}

The two effects linked to the BVF spikes, IGW trapping and enhanced radiative damping (in simulations), can be seen at play in the snapshots of Fig.~\ref{fig:mollweide_midMS_ZAMS}, which shows a comparison for the radial velocity $v_r$ at the same $r/R_\star$ locations for ZAMS (left) and midMS (right); specifically, we look at the flows at the convective boundary (top) as well as $0.02R_\star$ (middle) and $0.15R_\star$ (bottom) above it. The two stars have qualitatively similar flows at the CB (top row), with both showing plumes hitting the convective interface and generating IGWs. At $0.02R_\star$ above the CB (middle row), which for the midMS model roughly corresponds to the midway point within its BVF peak, both stars show significant IGW generation by the plumes observed at the CB (these can be seen as concentric circles in the figure). Looking past the BVF peak, at $0.15R_\star$ above the convective interface (bottom row) 
the ZAMS star still shows clear signs of both small and large scale IGW propagation throughout the $(\phi, \theta)$ domain; the midMS model, on the other hand, shows lower amplitude fluctuations, as both low-- and high-frequency waves have been heavily affected by the presence of the BVF spike with the former being particularly susceptible to the enhanced radiative damping and the latter remaining trapped in the BVF spike. This overall trend is likely to be reproduced in real stars where low frequency waves will be radiatively damped (though at a lower rate than these simulations) and high frequency waves will be trapped within the spike.

In order to visualize the frequencies excited in a particular model and how its spectrum evolves as a function of stellar radius, we create spectral heat maps integrated over all $l$ by sampling multiple longitudinal points across the stellar equator at all simulation timesteps; Fig.~\ref{fig:ZAMS_midMS_spectral_map} shows such spectral heat maps for the radial velocity $v_r$ as a function of both frequency $f$ and fractional stellar radius $r/R_\star$ for models 7ZR10 (labelled `Z', left) and 7mR10 (`m', right). The convection zones, which stretch up to $r/R_\star \approx 0.22$ and $r/R_\star \approx 0.13$, respectively, are characterized by the presence of all possible frequencies, thanks to convective motions being distributed over a large range of timescales, with $f \lesssim 30 \, \si{\micro Hz}$ (corresponding to timescales longer than $0.39\,\si{d}$) being the most dominant range. 

Within the radiation zone, the signal is instead constrained by the Brunt--V\"{a}is\"{a}l\"{a} frequency (white dashed lines), with a sharp drop for $f>N/2\pi$ as expected for IGWs. This behaviour can also be observed in the inset axes of the spectral heat map for model 7mR10 (right), which shows a zoomed in view around the BVF spike, where the signal is observed to closely mirror the $N/2\pi$ profile. Exempt from the restriction of the Brunt--V\"{a}is\"{a}l\"{a} frequency are the fundamental modes and gravity modes, the strong vertical features which are seen spanning significant radial portions of the RZ. Low-frequency waves appear strongly damped, with frequencies below approximately $10\, \si{\micro Hz}$ lacking throughout most of the RZ in both models. In our simulations this is largely due to enhanced thermal and viscous diffusion; however, one would expect a lack of low frequencies in real stars as well due to thermal diffusion.


\begin{figure}[t] 
    \includegraphics[width=1.\columnwidth]{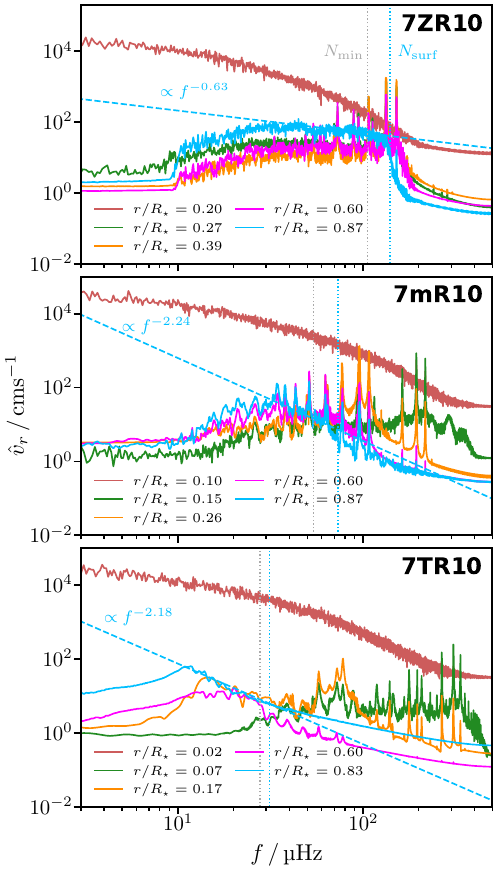}
    \caption{Frequency spectra for the radial velocity $v_r$ integrated over all $l$ for all three stellar ages (ZAMS, top; midMS, middle; TAMS, bottom) at multiple stellar radii. The blue line in each panel represents a power-law fit to the frequency spectrum profile taken closest to the domain's outer boundary of the model. The vertical lines in each panel represent the smallest BVF value within the RZ (grey, dashed lines) and the BVF value at the $r/R_\star$ location closest to the surface (light blue, dashed lines). }
	\label{fig:models-spectra}
\end{figure}

\begin{figure*}[t] 
	\includegraphics[width=\textwidth]{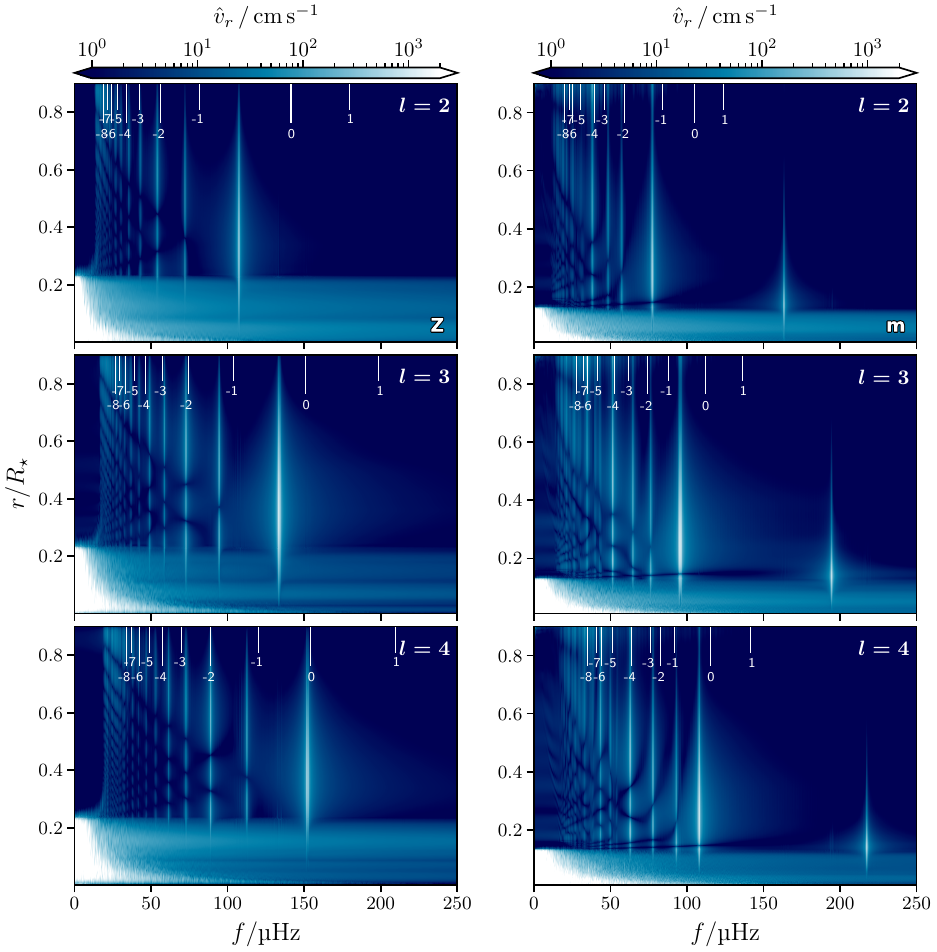}
    \caption{Frequency spectra of the radial velocity $v_r$ for the ZAMS (labelled `Z', left panels) and midMS (labelled `m', right panels) models for angular degrees $l=2$ (top panels), $l=3$ (middle panels) and $l=4$ (bottom panels). The white vertical lines (whose length has been varied to aid readability) are the mode frequencies for the relevant model and $l$ value as computed with GYRE. The modes are numbered according to the Eckart--Osaki--Scuflaire--Takata scheme, with negative numbers representing g modes, positive numbers are p modes and zero representing the f mode.}
	\label{fig:models-GYRE}
\end{figure*}

Figure~\ref{fig:models-spectra} shows integrated frequency spectrum profiles for $v_r$ for all three stellar ages, obtained by considering slices at fixed $r/R_\star$ from their respective spectral heat maps (such as those shown in Fig.~\ref{fig:ZAMS_midMS_spectral_map}). Each panel (top: ZAMS, middle: midMS, bottom: TAMS) shows spectrum profiles at five distinct fractional radial locations: in each case, the first profile is taken approximately $0.02 R_\star$ below the convective boundary (maroon lines), while the remaining four are spread throughout the radiation zone. The profiles within the radiation zones are taken at roughly $0.03 R_\star$ above the CB, which for the midMS and TAMS models falls within their BVF spikes (green lines); at the location where the Brunt--V\"{a}is\"{a}l\"{a} frequency reaches its maximum value, excluding BVF spikes (orange lines); at $r/R_\star = 0.60$ (magenta lines) and at approximately $0.03 R_\star$ below the outer edge of the domain (light blue lines). 
The minimum Brunt--V\"{a}is\"{a}l\"{a} frequency value throughout the radiation zone, $N_\mathrm{min}$, for each model is indicated by means of a vertical grey dotted line; additionally, the value of the Brunt--V\"{a}is\"{a}l\"{a} frequency at the chosen $r/R_\star$ closest to the surface, $N_\mathrm{surf}$, is indicated by light blue dashed lines, matching the colour of the profile obtained at the same location.

The most obvious difference between the models is the relative lack of evolution of the ZAMS spectrum as a function of $r/R_\star$ when compared to its older counterparts; the midMS and TAMS profiles are seen to vary considerably from within the BVF spike (green profiles) to the outer domain boundary, due to several effects such as filtering by the BVF spike, radiative damping and a more restrictive Brunt--V\"{a}is\"{a}l\"{a} frequency profile. While the profiles taken just above the CZ (green) are qualitatively similar across age, the phenomena mentioned above mean the surface signatures (light blue) of the three stellar ages differ greatly. 
The steepness of the spectra taken near the surface appears to increase from ZAMS, which presents a flatter spectrum with a power law exponent of roughly $-0.6$, to the older midMS and TAMS models whose exponent is found to be roughly $-2$. 
This effect appears to be a consequence of IGWs with frequencies approaching $N_\mathrm{surf}$ not reaching the outer boundary of the domain in the midMS and TAMS models, as evidenced by their gradual drop-offs in signal around $N_\mathrm{surf}$ (light blue dashed vertical lines) as opposed to the sharp decline observed for ZAMS. This is believed to be caused by large quantities of IGWs becoming evanescent in older stars due to the position of the turning point receding inwards with age; IGWs with frequencies approaching the local $N^2$ value at any particular $r/R_\star$ would be the most heavily affected by the turning point receding inward and would become evanescent further away from the stellar surface than waves with lower frequencies, as suggested by Eq.~\ref{eq:turning-point}. 
The steepness of all models considered is however consistent with other 2D and 3D simulations, such as \cite{RogersMacGregor2010, Rogersetal2013, Alvanetal2014, Augustsonetal2016, Edelmannetal2019}, who observed spectra with power law exponents roughly between $-1$ and $-3$. 

Both Fig.~\ref{fig:ZAMS_midMS_spectral_map} and Fig.~\ref{fig:models-spectra} exhibit gravity and fundamental modes in the form of vertical features; a more thorough analysis on their nature and the accuracy of their frequencies is carried out in Fig.~\ref{fig:models-GYRE}, which shows the frequency spectra of the radial velocity $v_r$ for all radii for the ZAMS (labelled `Z', left column) and midMS (labelled `m', right column) models for individual angular degrees: $l=2$ (top panels), $l=3$ (middle panels) and $l=4$ (bottom panels). This allows us to disentangle the contributions from different $l$ values and obtain a clearer view of the resonant coherent modes within the stellar cavity. The nature of the modes can be established by the number of radial nodes they present; the modes without any nodes are f-modes, or fundamental modes, while modes exhibiting an increasing amount of radial nodes with decreasing frequency are g-modes, or gravity modes. The modes observed within each panel are compared to their predicted frequencies obtained from the stellar oscillation code GYRE \citep{TownsendTeitler2013, Townsendetal2018}, which are represented by means of vertical white lines and labelled according to the Eckart--Osaki--Scuflaire--Takata scheme \citep[see e.g.,][]{Aertsetal2010}; this means the f-mode is indicated by 0, g-modes are represented by negative numbers and p-modes by positive numbers. The latter are however not picked up in our simulations due to the anelastic nature of the code\footnote{For this reason, p-modes beyond p$_1$ are not shown}. 

For the ZAMS model there is good agreement between the simulation and GYRE concerning the g-modes with g$_2$, g$_3$ and g$_4$ matching particularly well, especially for $l=2$ and $l=3$. The level of agreement for the f-modes varies more noticeably across $l$; for $l=2$ the observed f-mode sits roughly halfway between the frequencies predicted by GYRE for the g$_1$ and the f modes, but for $l=4$ the simulation and the data from GYRE match closely. This improved agreement for higher values of $l$ is actually in contrast with the analogous analysis from \cite{Edelmannetal2019}, where the level of agreement between GYRE and the simulation data for the fundamental modes was found to instead decrease with increasing $l$. 

The agreement with GYRE is however more erratic for the midMS model. The high frequency f mode (clearly located within the BVF spike at $f>150\,\si{\micro Hz}$) appears completely mismatched with the frequencies expected by GYRE and although the g modes match well for $l=2$ and $l=3$, their agreement worsens for $l=4$ where the simulation overestimates the frequencies of modes g$_1$ and g$_2$.
The reason for this inconsistent agreement between our midMS model and GYRE is not clear, but two likely explanations have been singled out (as suggested by R.~H.~D. Townsend in a private conversation). One option is that it could be caused by an eigenfrequency distortion in GYRE due to the software not properly dealing with mass conservation within the BVF spike. Alternatively, it could be a natural limitation of our anelastic code when dealing with older stars: the emergence of the BVF spike in the midMS model (and TAMS models) increases the maximum frequency allowed for the propagation of gravity waves; this has the consequence of overlapping the frequency ranges for gravity waves and sound waves (which is demonstrated by our midMS simulation finding the f modes at frequencies higher than the GYRE p$_1$ frequency). Any modes falling within this overlap region would take a `mixed mode' character \citep{Unnobook1989}, and the anelastic nature of our code might struggle to properly interpret and reproduce these frequencies.

\begin{figure*}
    \centering
    \includegraphics[width=\textwidth]{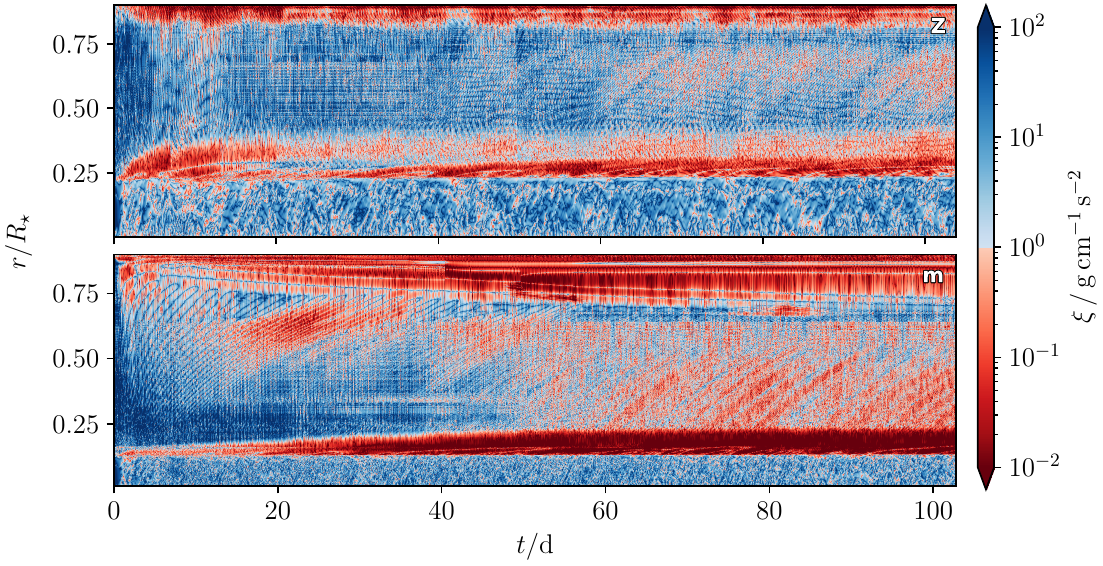}
    \caption{Temporal evolution of the ratio between the divergences of the Reynolds and viscous stresses, $\xi$, as a function of fractional radius for the ZAMS (Z, top) and midMS (m, bottom) models. $\xi > 1$ values (blue) correspond to the Reynolds stress dominating, while for $\xi < 1$ (red) viscous fluxes are stronger.}
    \label{fig:ZAMS-midMS-reyn-visc-ratio}
\end{figure*}

\subsection{Angular momentum transport} \label{sec:results-angularmomentum}
IGWs play an important role in the transfer of angular momentum within stellar interiors thanks to their ability to propagate and dissipate within radiation zones; this is particularly true for massive stars, such as those studied here, where the stellar density stratification causes some IGW amplitudes to grow during their propagation towards the stellar surface.

Transport of angular momentum via IGWs is however only possible when the waves are attenuated, as demonstrated by \cite{EliassenPalm1960}. There exist three main mechanisms that can dissipate IGWs within stellar interiors: radiative damping, the formation of critical layers and non-linear wave breaking due to large amplitude. 
Radiative damping (Eq.~\ref{eq:radiative-diff}--\ref{eq:radiative-damping-length}) affects wave amplitudes non-linearly and is very sensitive to wave frequency, being at its most effective for low frequencies. Critical layers are the product of a co-rotation resonance, which is achieved when the Doppler-shifted frequency of a wave

\begin{equation}
    \omega(r) = \omega_\mathrm{gen} - m \left[\Omega(r) - \Omega_\mathrm{gen}\right] 
\end{equation}
becomes zero, where $\omega_\mathrm{gen}$ is the frequency of such a wave at the generation radius (ie. at the CB), and $\Omega_\mathrm{gen}$ and $\Omega(r)$ are the angular velocities of the flow at the CB and at radius $r$.  
Interactions between IGWs and critical layers can either result in wave attenuation or in the IGWs being reflected/transmitted, depending on the value of the Richardson number at the critical layer \citep{Alvanetal2013}. While our models are started with a solid body rotation rate of either $10^{-6}$ or $10^{-5} \, \si{rad \, s^{-1}}$ (see Table~\ref{table:models}), they quickly develop a weak differential rotation between the convective and radiative regions and near the simulated surface; given the strength of such differential rotation, we would expect critical layers to occur in these regions for waves possessing frequencies $\omega_\mathrm{gen} \sim m \times 10^{-7} \, \si{rad \, s^{-1}}$. 
Lastly, non-linear wave breaking can occur either through convective instability or a Kelvin-Helmholtz instability; the former is restricted to regions where $N \sim 0$, such as the interfaces between convective and radiative zones, meaning a Kelvin-Helmholtz instability is the most likely source of wave breaking within the bulk of stellar radiative zones. Kelvin-Helmholtz instabilities develop when the destabilizing shearing motions between adjacent layers of a stratified shear flow prevail over the buoyancy; numerically, this is controlled by the Richardson number

\begin{equation}
    \mathrm{Ri} = \frac{N^2}{k_r^2 u_h^2},
\end{equation}
with $\mathrm{Ri}\lesssim 1$ yielding a Kelvin-Helmholtz instability. For IGWs, this instability criterion can be expressed as $\epsilon = k_h u_h/\omega \gtrsim 1$  \citep{Press1981}. It is unclear whether this instability occurs due to IGW in stellar radiative interiors.  While these 3D simulations do not demonstrate this instability, it was seen in previous 2D simulations.  The difference being due to slightly larger forcing and lower damping in 2D compared to 3D.  Given it is unclear what combination of forcing/damping might accurately represent the radiative interior of a star (if any), the jury is still out.  However, work by \cite{LeSauxetal2023} indicates that nonlinearity of the waves is likely only in limited circumstances. 


Whenever an IGW is attenuated, it transfers angular momentum to the mean flow \citep{EliassenPalm1960}. The dynamical interaction between the waves and the mean flow are complex and highly non-linear, being dependent on the properties of both
. Neglecting meridional circulation, the evolution of angular momentum transport via IGWs is described by
\begin{equation} \label{eq:ang-mom} \begin{multlined}
    \rho \partial_t \left( r^2 \langle{\Omega}\rangle\right) = \\ 
    - \frac{1}{r^2}\partial_r \left(r^2 \rho r \langle{\sin \theta v_r v_\phi}\rangle\right) + \frac{1}{r^2} \partial_r \left(\rho \nu r^4 \partial_r \langle{\Omega}\rangle\right) ,
\end{multlined}\end{equation}
where angle brackets $\langle \cdot \rangle$ represent horizontal averages and $\langle{\Omega}\rangle = \int_0^\pi \Omega \sin^3 \theta \mathrm{d}\theta / \int_0^\pi \sin^3 \theta \mathrm{d}\theta$ \citep{Zahn1992, MaederZahn1998, MathisZahn2004}. As seen from Eq.~\ref{eq:ang-mom}, the evolution of the mean flow is controlled by the divergence of the horizontally averaged Reynolds stress, $F_\mathrm{Rey} = \rho r \langle{\sin \theta v_r v_\phi}\rangle$, and by the divergence of the viscous flux. Comparing the magnitudes of these two terms allows us to establish which term is dominant and, consequently, understand the dynamics of the stellar interiors. With that aim in mind, we define the ratio between the divergences of the Reynolds stress and the viscous flux as

\begin{equation}
    \xi = \frac{\partial_r \left(r^2 \rho r \langle{\sin \theta v_r v_\phi}\rangle\right)}{\partial_r \left(\rho \nu r^4 \partial_r \langle{\Omega}\rangle\right)}.
\end{equation}

\begin{figure*}[th]
    \centering
    \subfloat{
        \includegraphics[width=\columnwidth]{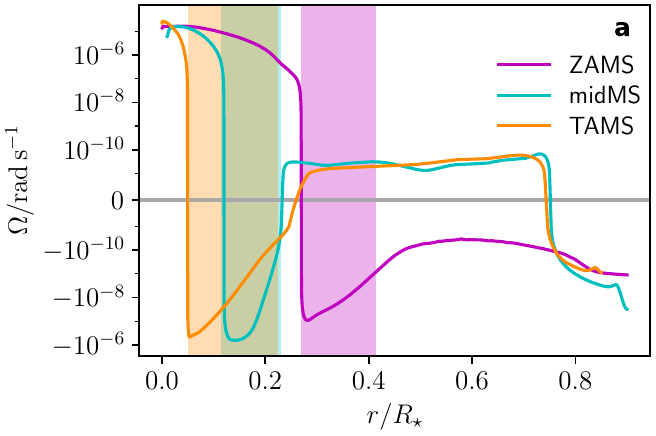}
        \label{fig:Omega_profiles}
        }
    \hspace*{2em}
    \subfloat{
        \includegraphics[width=\columnwidth]{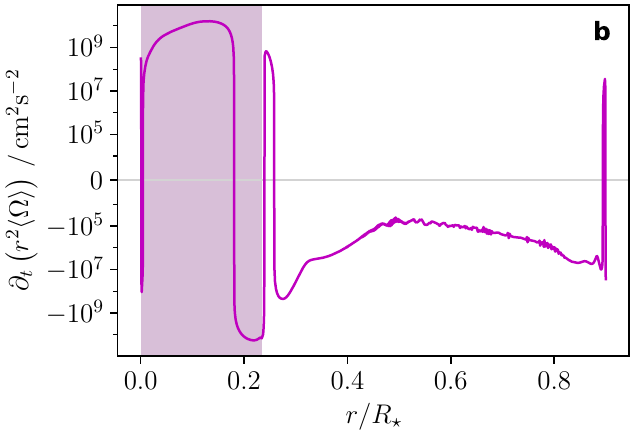}
        \label{fig:j_profile_ZAMS}
    }
    \caption{\textbf{\textit{(a):}} Radial profiles of the time-averaged angular velocity perturbation $\Omega$ from the background rotation rate $\Omega_0$ for ZAMS (magenta), midMS (cyan) and TAMS (orange). The colour-coded shaded regions indicate the depths of the shear flows located directly above the CZ of their respective models. \textbf{\textit{(b):}} Radial profile for the time-averaged rate of specific angular momentum transport, $\partial_t \left(r^2 \langle \Omega \rangle \right)$ for the ZAMS model; the shaded area illustrates the location of the CZ.}
\end{figure*}

The evolution of $\xi$ over time as a function of radius is shown in Fig.~\ref{fig:ZAMS-midMS-reyn-visc-ratio} for both ZAMS (Z, left) and midMS (m, right) models. The results from the TAMS model are not shown because the low amplitudes of IGWs, due to the effects described above, mean this model is viscously dominated everywhere within the radiative region. In both ZAMS and midMS models, the convection zones are dominated by Reynolds stresses (blue); immediately above the convective boundary, both models present an area dominated by viscous flux (red), due to the strong shear flow which develops between the convective and radiative regions. This shear flow partly overlap with the overshooting regions of the two models, where the value of the Brunt--V\"{a}is\"{a}l\"{a} frequency is low, causing the overshoot regions to be potentially susceptible to shear instability or convective instability, the latter of which could result in non-linear wave breaking as mentioned earlier. Additionally, both stars exhibit another shear flow at the outer boundary of their domain, with the midMS model featuring a shear flow that is seemingly more radially extended.
Above the first shear flow some differences between the two models can be observed: the bulk of the radiation zone in the ZAMS model is dominated by Reynolds stresses, indicating the prominence of IGWs in the dynamics of this region at this age; the same region for the midMS model, on the other hand, exhibits both Reynolds stresses and viscous fluxes alternating in a seemingly stochastic nature. This disparity in the importance of IGWs within the bulk of the radiation zones is due to the same effects described in Section~\ref{sec:results-propagation}, i.e. trapping of waves within the BVF spike and waves becoming evanescent at a lower fractional radius within the radiative zone.  While we expect these effects to be relevant in stars, they are exacerbated in these simulations due to the enhanced viscosity. Hence, while we expect older stars to have lower Reynolds stress amplitudes, we also expect them to have significantly lower viscous flux.  Which is dominant in actual stars is unclear. 

The conclusions drawn from the $\xi$ plots are reinforced by Fig.~\ref{fig:Omega_profiles}, which shows the time-averaged radial profiles of the angular velocity fluctuation $\Omega$ for all three stellar ages (models 7ZR10, 7mR10, 7TR10). As mentioned previously, these three models are all set up with an uniform stellar rotation of $\overline{\Omega}=10^{-5} \, \si{rad \, s^{-1}}$ (as indicated in Table~\ref{table:models}), but are seen to develop differentially rotating radial profiles. In all models the convective cores exhibit a sizeable positive angular velocity fluctuation, while the shear flows above them are rotating slower than the background value. The depths of these shear flows are highlighted by the colour-coded shaded areas; all models present similarly deep shear layers, with their radial extents ranging from $0.116 R_\star$ (midMS) to $0.174 R_\star$ (TAMS). 
This results in the central part of the radiative zone rotating slower than the background ration for the ZAMS model, while in the midMS and TAMS models the same region rotates slightly faster than the initial uniform rotation.
Near the outer boundary we observe another band of negative $\Omega$ in both models, representing the surface shear flows observed in the $\xi$ plot, which are likely maintained by IGWs.

Finally, Fig.~\ref{fig:j_profile_ZAMS} shows the radial profile for the rate of specific angular momentum transport, $\partial_t \left(r^2 \langle \Omega \rangle\right)$, as obtained from Eq.~\ref{eq:ang-mom} for the ZAMS model. This profile shows strong similarities to the $\Omega$ profile in Fig.~\ref{fig:Omega_profiles}, with values of $\lvert \partial_t \left(r^2 \langle \Omega \rangle \right) \rvert \sim 10^{10} \, \mathrm{cm^{2} \, s^{-2}}$ recorded in the core and $\sim 10^{6} \, \mathrm{cm^{2} \, s^{-2}}$ in the bulk of the RZ. This latter value implies that IGWs would be able to alter the rotation rate of the radiative zone in a timescale

\begin{equation}
    t_\mathrm{IGW} = \frac{R_\mathrm{RZ}^2 \Omega}{\partial_t \left(r^2 \langle \Omega \rangle \right)} \sim 8\times 10^8 \, \mathrm{yrs},
\end{equation}
where $R_\mathrm{RZ}$ is the radial extent of the radiative zone. This timescale is significantly longer than the one estimated by \cite{Fulleretal2014} for solar-type stars, giants and sub-giants. It is however important to highlight that the models presented in this work 
self-consistently account for both prograde and retrograde IGWs and autonomously develop differential rotation as a result of  their IGW propagation properties.

\section{Discussion and Conclusions} \label{sec:conclusions}
We present 3D simulations of a 7M$_\odot$ star for three distinct stellar ages (ZAMS, midMS and TAMS), including both their convective cores and a large fraction of the radiative zones, extending up to $90\%$ of the total stellar radius. 

Generation spectra taken both just below and above the convective boundary do not appear dominated by a single frequency as might be expected for a spectrum strongly influenced by convective eddies; furthermore, the generation spectra taken below the CB are seen to peak at frequencies roughly one order of magnitude higher than their respective turnover frequencies.

An analysis into overshooting plumes within our simulations reveals two distinct overshooting layers above the convective boundary, with the deeper layer becoming increasingly more difficult to reach as the star ages; this is a consequence of its convective interface becoming stiffer against convective motions due to the emergence of the spike in the Brunt--V\"{a}is\"{a}l\"{a} frequency profile. This also explains the average plume overshooting length decreasing with stellar age. 

A direct comparison with several theoretical prescriptions, including both eddy-dominated and plume-dominated profiles, indicates our generation spectra are dominated by plumes, being particularly well described by the profiles outlined in \cite{Pinconetal2016, Pinconetal2017}. Broken power laws also fit the spectra adequately well
, with exponents similar to those found in the 2D numerical work by \cite{Rogersetal2013}, suggesting the properties of the generation spectra are largely unaffected by dimensionality.



An analysis of frequency spectrum profiles across the radiation zones of the three stars shows that the spectra of the midMS and TAMS models evolve significantly throughout their radiation zones, while the ZAMS spectrum remains largely unchanged. This is caused in part by the presence of a BVF spike in older stars, which traps high frequency waves, and by the stellar density stratification affecting IGW propagation; this causes waves to become evanescent in the radiative zone, a mechanism that is particularly strong in older stars. Strong radiative damping in the BVF spikes also plays a role, although this effect would not be significant in real stars.
Given our simulations require significantly enhanced viscosity and thermal diffusivity values, we are confident that the two non-damping effects mentioned above would also be relevant in real stars.

Furthermore, the steepness of the profiles at the surface is found to increase from ZAMS to the older midMS and TAMS models; this is believed to be largely caused by the turning point location receding inwards with age, which prevents IGWs with frequencies approaching the local BVF from reaching the outer boundary of the star. 
This behaviour is also seen in observations of stochastic low-frequency variability, where the spectral power of near-ZAMS stars is distributed over a larger frequency range compared to older near-TAMS stars in terms of their respective observed characteristic frequencies \citep{Bowmanetal2019}.

The observed f and g modes in the frequency spectra for the ZAMS and midMS models were also compared to the mode frequencies predicted by the stellar oscillation code GYRE; the g modes appear well matched in both models, with the f mode in the ZAMS star also showing satisfactory agreement; the midMS f mode appears however poorly matched with the GYRE predicted frequencies, likely a result of the overlapping frequency ranges for g modes and p modes (not resolved by our anelastic code) giving rise to `mixed modes', which are poorly captured by our code. 

As a consequence of older stars being more scarcely populated with waves, the dominance of the Reynolds stress within the radiation zone wanes with stellar age. All models exhibit two shear flows: one located directly above the CZ, and one at the outer boundary. Despite all models were initialized with a uniform rotation rate, they all develop similar weakly differential rotation profiles, with the core spinning quicker than the background rate while the two shear flows rotate slower than it. From radial profiles of the angular velocity fluctuation it is also possible to extract the depths of the shear layers located directly above the CZs, which were found to be consistent across stellar age. 

Future work will concentrate on employing more realistic luminosity and thermal diffusivity values to slowly bridge the gap between real stellar interiors and simulations. Furthermore, access to more computational resources would provide us with the opportunity of adding sub-surface convection zones in our simulations, improving our analysis of IGW dynamics in evolved stars.

\section*{Acknowledgements}
We acknowledge support from STFC grant ST/L005549/1 and NASA grant NNX17AB92G. Computing was carried out on Pleiades at NASA Ames, Rocket High Performance Computing service at Newcastle University and DiRAC Data Intensive service at Leicester, operated by the University of Leicester IT Services, which forms part of the STFC DiRAC HPC Facility (\url{www.dirac.ac.uk}), funded by BEIS capital funding via STFC capital grants ST/K000373/1 and ST/R002363/1 and STFC DiRAC Operations grant ST/R001014/1. PVFE was supported by the U.S. Department of Energy through the Los Alamos National Laboratory (LANL). LANL is operated by Triad National Security, LLC, for the National Nuclear Security Administration of the U.S. Department of Energy (Contract No. 89233218CNA000001). This paper was assigned a document release number LA-UR-22-28600. The authors thank R.~H.~D. Townsend for helpful comments. 





\bibliography{7Ms_3D_IGW_aastex}{}
\bibliographystyle{aasjournal}

\end{document}